\documentclass[longauth]{aa}
\bibpunct{(}{)}{;}{a}{}{,} % to follow the A&A style
\usepackage{graphicx}
\usepackage{subfig}
\usepackage{hyperref}
\usepackage{comment}
\usepackage{xcolor}
\usepackage[normalem]{ulem}

\title{Evidence for magnetic boundary layer accretion in RU~Lup. A~spectrophotometric analysis\thanks{Based on observations collected at the European Southern Observatory under ESO programs 106.20Z8.003 and 106.20Z8.007.}}
\author{A. Armeni\inst{\ref{inst1}} \and B. Stelzer\inst{\ref{inst1}} \and A. Frasca\inst{\ref{inst2}} \and C.~F. Manara\inst{\ref{inst3}} \and F.~M. Walter\inst{\ref{inst4}} \and J.~M. Alcalá\inst{\ref{inst5}} \and P.~C. Schneider\inst{\ref{inst6}} \and A.~Sicilia-Aguilar\inst{\ref{inst7}} J.~Campbell-White\inst{\ref{inst3}} \and E.~Fiorellino\inst{\ref{inst5}} \and J.~F.~Gameiro\inst{\ref{inst8},\ref{inst9}} \and M.~Gangi\inst{\ref{inst10},\ref{inst11}} }

\institute{
    Institut für Astronomie und Astrophysik, Eberhard Karls Universität Tübingen, Sand 1, 72076 Tübingen, Germany \email{armeni@astro.uni-tuebingen.de} 
    \label{inst1}
    \and
    INAF – Osservatorio Astrofisico di Catania, via S. Sofia 78, 95123 Catania, Italy
    \label{inst2}
    \and
    European Southern Observatory, Karl-Schwarzschild-Strasse 2, 85748 Garching bei München, Germany
    \label{inst3}
    \and
    Department of Physics \& Astronomy, Stony Brook University,
    Stony Brook NY 11794-3800, USA
    \label{inst4}
    \and
    INAF – Osservatorio Astronomico di Capodimonte, via Moiariello 16, 80131 Napoli, Italy
    \label{inst5}
    \and
    Hamburger Sternwarte, Gojenbergsweg 112, D-21029, Hamburg, Germany
    \label{inst6}
    \and
    SUPA, School of Science and Engineering, University of Dundee, Nethergate, DD1 4HN, Dundee, UK
    \label{inst7}
    \and
    Instituto de Astrofísica e Ciências do Espaço, Universidade do Porto, CAUP, Rua das Estrelas, P-4150-762 Porto, Portugal
    \label{inst8}
    \and
    Departamento de Física e Astronomia, Faculdade de Ciências, Universidade do Porto, Rua do Campo Alegre 687, P-4169-007 Porto, Portugal
    \label{inst9}
    \and
    INAF – Osservatorio Astronomico di Roma, via Frascati 33, 00078 Monte Porzio Catone, Italy
    \label{inst10}
    \and
    ASI, Italian Space Agency, via del Politecnico snc, 00133 Rome, Italy
    \label{inst11}
}

\date{14/02/2024}

\abstract {It is well established that classical T Tauri stars accrete material from a circumstellar disk through magnetic fields. However, the physics regulating the processes in the inner (0.1 AU) disk is still not well understood.}{Our aim is to characterize the accretion process of the classical T Tauri Star RU~Lup.}
{Optical high-resolution spectroscopic observations with CHIRON and ESPRESSO were obtained simultaneously with photometric data from AAVSO and TESS.}
{We detected a periodic modulation in the narrow component of the \ion{He}{i} 5876 line with a period that is compatible with the stellar rotation period, indicating the presence of a compact region on the stellar surface that we identified as the footprint of the accretion shock. 
We show that this region is responsible for the veiling spectrum, which is made up of a continuum component plus narrow line emission that fills in the photospheric lines.
An analysis of the high-cadence TESS light curve reveals quasi-periodic oscillations on timescales shorter than the stellar rotation period, suggesting that the accretion disk in RU~Lup extends inward of the corotation radius, with a truncation radius at $\sim 2 ~ R_{\star}$. This is compatible with predictions from three-dimensional magnetohydrodynamic models of accretion through a magnetic boundary layer (MBL). In this scenario, the photometric variability of RU~Lup is produced by a nonstationary hot spot on the stellar surface that rotates with the Keplerian period at the truncation radius.
We also qualitatively discuss how more complex hot spot shapes may generate the same variability pattern.
The analysis of the broad components of selected emission lines reveals the existence of a non-axisymmetric, temperature-stratified flow around the star, in which the gas leaves the accretion disk at the truncation radius and accretes onto the star channeled by the magnetic field lines.
The unusually rich metallic emission line spectrum of RU~Lup might be characteristic of the MBL regime of accretion.}{Our extensive multiwavelength database of RU~Lup reveals many similarities to predictions from the scenario of accretion through a magnetic boundary layer. Alternative explanations would require the existence of a hot spot with a complex shape, perhaps made of two brighter knots, or a warped structure in the inner disk.}


\begin{document}
    \titlerunning{MBL accretion in RU Lup}
    \authorrunning{A. Armeni et al.}
    \maketitle
    \section{Introduction}
    Classical T Tauri stars (CTTSs) are young ($\sim 1-10$ Myr), low-mass ($< 2 \, M_{\odot}$) objects that accrete material from a circumstellar disk \citep{Hartmann+2016}. Their strong magnetic fields truncate the disk at a few stellar radii (typically 5~$R_{\star}$, \citealt{Hartmann+2016}). 
    The current paradigm for the interaction between the disk and the star is magnetospheric accretion \citep{Bouvier+2007}, in which the material free-falls onto the star following the magnetic field lines. 
    
    The rich emission line spectrum is one of the defining characteristics of these systems \citep{Joy1945, Herbig1962}. The optical spectrum of CTTSs displays strong and broad (with a full width at zero intensity $\gtrsim 200 ~ \rm{km~s^{-1}}$) permitted emission lines, such as the Balmer, \ion{He}{i}, and \ion{Ca}{ii} H \& K lines.
    % as well as forbidden lines such as the [\ion{O}{i}] 6300. 
    Sometimes other metallic species such as \ion{Na}{i}, \ion{Mg}{i}, \ion{Ca}{i}, \ion{Ca}{ii}, \ion{Fe}{i}, and \ion{Fe}{ii} are observed in emission, especially during epochs of an increased accretion rate \citep[e.g.,][]{Sicilia-Aguilar+2012, Armeni+2023}.
    Permitted lines are thought to be formed in different structures around the star. The observed supersonic velocities, roughly consistent with free-fall velocities, suggest that the broad component (BC) of permitted emission lines is formed in the magnetospheric accretion flow \citep{Hartmann+2016}. 
    % On the other hand, forbidden lines indicate the presence of mass loss in the form of jets and/or disk winds \citep{Bally2016, Banzatti+2019, RayFerreira2021}.
    Many studies have shown that the Balmer and other emission line profiles and variability are compatible with this scenario \citep[e.g.,][]{Muzerolle+1998, Alencar+2012, Bouvier+2007b} and that their luminosity is related to the accretion rate of the system \citep{Alcala+2017}. 
    The helium and metallic lines are also useful tracers of the dynamics of the accretion flow \citep{Beristain+1998, Beristain+2001}. These lines typically show a narrow component (NC) superposed on the BC.
    The NC is formed in the footprint of the magnetic field on the stellar surface - that is, the post-shock region - and it is rotationally modulated with the stellar rotation period \citep{Sicilia-Aguilar+2015, Campbell-White+2021}. 
    % The BC originates instead in the circumstellar environment.
    The different behavior between the BC of the \ion{He}{i} and metallic lines and their different excitation conditions suggest the presence of temperature and density gradients in the accretion flow \citep{Armeni+2023}. In particular, the \ion{He}{i} lines require high-energy conditions that can be achieved in the pre-shock region, which is irradiated by the X-rays from the shock \citep{Hartmann+2016}. Under such conditions, species such as Na, Ca, and Fe are expected to be highly ionized. For this reason, the low-ionized metallic lines are thought to originate closer to the disk \citep{Sicilia-Aguilar+2012, Sicilia-Aguilar+2023, Armeni+2023}, where the dilution factor of the radiation from the shock is lower \citep{Azevedo+2006}.

    Another observational feature in the optical spectrum of CTTSs is the presence of a continuum excess flux relative to the photospheric spectrum, which results from the accretion shock at the stellar surface \citep{CalvetGullbring1998}. This excess makes the photospheric absorption lines appear less deep in normalized spectra, an effect known as veiling \citep{Hartigan+1989}. The veiling fraction (VF) is defined as the ratio of the excess flux, $F_{\rm{acc}}$, due to accretion relative to the photospheric flux, $F_{\rm{phot}}$; that is, $\rm{VF}_{\lambda} = F_{\rm{acc}}(\lambda)/F_{phot}(\lambda)$. 
    % Therefore, comparison of observed CTTSs spectra with photospheric standards of non-accreting objects allows to discern the magnitude of this excess. 
    % In the magnetospheric accretion scenario, the veiling is expected to correlate with the brightness of the system. However, \citet{Gahm+2008} found that this is not the case for extremely veiled (VF $\gtrsim 2$) CTTSs. They interpreted this lack of correlation as due to the effect of line filling emission in the absorption lines, that dilutes the photospheric spectrum without any changes in the continuum emission from shocked regions.
    \citet{Gahm+2008} showed that for extremely veiled (VF~$\gtrsim~2$) CTTSs another component contributes to the veiling: line filling emission in the absorption lines, which dilutes the photospheric spectrum without any changes in the continuum emission from shocked regions.
    
    Three-dimensional (3D) magnetohydrodynamic (MHD) simulations showed that the interaction between the star and the disk is complex \citep{RomanovaOwocki2015}. It has been shown that CTTSs may accrete in either a stable or an unstable regime \citep{Romanova+2003, Romanova+2004, KulkarniRomanova2008, Pantolmos+2020}. 
    In the stable regime, the disk matter is channeled in funnel streams by the magnetic fields, forming polar hot spots on the stellar surface. In the Rayleigh-Taylor (RT) unstable regime, the matter penetrates through the magnetosphere in tongues. This leads to the formation of several chaotic hot spots close to the stellar equator. 
    The light curves are expected to be periodic with the stellar rotation period ($P_{\star}$) in the stable regime and stochastic in the unstable regime. % with the absence of $P_{\star}$ in the power spectrum of the light curves, 
    
    According to the simulations, the transition between the stable and unstable regimes depends on the ratio between the magnetospheric truncation radius, $R_{\rm{T}}$, and the corotation radius, $R_{\rm{co}}$.
    The corotation radius is the radius at which the Keplerian angular velocity of the disk matches the angular velocity of the star. % $R_{\rm{co}} = \left(G M_{\star}\right)^{1/3} \left( P_{\star}/2\pi \right)^{2/3}$, where $G$ is the gravitational constant and $M_{\star}$ is the mass of the star.
    % The truncation radius is defined as the radius in the disk where the magnetic field pressure ($P_{\rm B}$) is equal to the disk ram pressure ($P_{\rm D}$). Since $P_{\rm B} \propto B_{\star}^2$ and $P_{\rm D} \propto \dot{M}_{\rm acc}$, where $B_{\star}$ is the strength of the stellar dipolar magnetic field and $\dot{M}_{\rm acc}$ is the accretion rate in the disk, $R_{\rm{T}}$ depends on the ratio $B_{\star}^2/\dot{M}_{\rm acc}$. 
    The truncation radius is defined as the radius where the magnetic field pressure is equal to the ram pressure in the disk.
    By fitting the results of a series of 3D MHD simulations, \citet{KulkarniRomanova2013} showed that 
    \begin{equation}
        \frac{R_{\rm T}}{R_{\star}} = 1.06 \left( \frac{B_{\star}^4 R_{\star}^5}{\dot{M}_{\rm acc}^2 G M_{\star}} \right)^{1/10},
        \label{RT_Romanova}
    \end{equation}
    where $R_{\star}$ is the stellar radius, $B_{\star}$ is the strength of the stellar dipolar magnetic field at the stellar surface, $\dot{M}_{\rm acc}$ is the mass accretion rate from the disk, $G$ is the gravitational constant, and $M_{\star}$ is the stellar mass. All parameters are in cgs units.
    % Assuming spherical accretion and a Keplerian velocity in the disk, it can be shown that $R_{\rm{T}} \propto (B_{\star}^2/\dot{M}_{\rm acc})^{2/7}$, where $B_{\star}$ is the strength of the dipolar magnetic field of the star and $\dot{M}_{\rm acc}$ is the accretion rate in the disk \citep[e.g.,][]{Bouvier+2007}.
    Accretion is unstable if $R_{\rm{T}}/R_{\rm{co}} \lesssim 0.71$ and stable otherwise \citep{Blinova+2016}. % Therefore, systems with high $\dot{M}_{\rm acc}$ ($\gtrsim 2 \times 10^{-8} ~ M_{\odot}~\rm{yr^{-1}}$ for typical CTTSs parameters, \citealt{KulkarniRomanova2008}) are expected to accrete in the unstable regime.

    % When accretion is strongly controlled by the RT instability, the hot spots are entirely created by tongues. \citet{KulkarniRomanova2009} showed that the rotational period of the tongues and the spots is associated to the dynamical timescale of the inner disk, that is, the Keplerian rotation period $P_{\rm T}$ at the truncation radius. Although the light curves in this regime are expected to be stochastic due to the intrinsic variability caused by the RT instability, \citet{KulkarniRomanova2009} suggested that it might be possible to observe quasi-periodic oscillations (QPOs) at a period close to $P_{\rm T}$ in the power spectrum of the light curves.
    % , related to $R_{\rm{T}}$ by $R_{\rm{T}} = \left(G M_{\star}\right)^{1/3} \left(P_{\rm T}/2\pi \right)^{2/3}$.    
    
    \citet{RomanovaKulkarni2009} showed that in systems with small magnetospheres, unstable accretion proceeds through a magnetic boundary layer (MBL). In this regime, two ordered streams are formed. These tongues of matter and the resulting hot spots rotate with the Keplerian period, $P_{\rm T}$, at the truncation radius. This leads to the observation of quasi-periodic oscillations (QPOs) at $P_{\rm T}$ in the light curves.
    % that is, when either $B_{\star}$ is low or $\dot{M}_{\rm acc}$ is high, 
    % These QPOs can be either at $P_{\rm T}$ or $2P_{\rm T}$, depending on the geometry of the system. 
    In their simulations, the authors observed a correlation between the size of the magnetosphere and the period of the QPOs. When the magnetosphere is smaller, the detected period decreases. 
    % They suggested that variations in $\dot{M}_{\rm acc}$, which in turn change the position of $R_{\rm{T}}$, might explain the drifting of the QPO frequency observed in accreting millisecond pulsars \citep[e.g.,][]{vanderKlis2000}. 
    \citet{Blinova+2016} further investigated the conditions for the development of this regime, showing that it occurs for systems with $R_{\rm T} \lesssim 4.2~R_{\star}$ when $R_{\rm T}/R_{\rm co} \lesssim 0.59$.

    This work focuses on RU~Lup (Sz 83), a young K7 star \citep{Alcala+2017} located in the Lupus cloud at a distance of $158.9 \pm 0.7$~pc \citep{GaiaDR3}. RU~Lup is a monitoring target within the \textit{Hubble UV Legacy Library of Young Stars as Essential Standards}, \citep[ULLYSES,][]{RomanDuval+2020, Espaillat+2022} survey for CTTSs. 
    \citet{Manara+2023} reported a value of $\dot{M}_{\rm acc} = 10^{-7} ~ M_{\odot}~\rm{yr^{-1}}$ for RU~Lup, making it the strongest accretor in Lupus. This star is also very active, with an accretion rate that can vary up to a factor of 2 on a timescale of weeks \citep{Stock+2022}. \citet{Wendeborn+2024_1} show that the accretion rate reached median values of $1.7 \times 10^{-7} ~ M_{\odot}~\rm{yr^{-1}}$ in 2021, and it dropped considerably one year later, with a median of $6 \times 10^{-8} ~ M_{\odot}~\rm{yr^{-1}}$.
    \citet{Stempels+2007} derived $P_{\star} = 3.71$~d by analyzing the radial velocity of the photospheric absorption lines. Many works have focused on the photometric variability of RU~Lup, but failed to recover any stable periodicity \citep{Percy+2010, Siwak+2016, Wendeborn+2024_2}.
    
    \citet{Herczeg+2005} estimated $A_{\rm v} \sim 0.07$~mag for RU~Lup. This value is compatible with its pole-on orientation, obtained from both spectroscopy \citep{Stempels+2007} and interferometry \citep{GravityColl+2021}. This allows us to study the accretion process and its variability without the contribution of variable extinction due to circumstellar dust in our line of sight. 
    The optical spectrum of RU~Lup comprises a plethora of emission lines from metallic species. 
    % The line spectrum is richer than the bursting state spectrum of HM~Lup \citep{Armeni+2023} and similar to the outburst spectrum of EX~Lup \citep{Sicilia-Aguilar+2015} (see Sect.~\ref{spectroscopic_variability}).

    These characteristics make RU~Lup suitable for studying the magnetospheric accretion process in a likely RT-unstable regime \citep[given its high $\dot{M}_{\rm acc}$,][]{Stock+2022} by comparing photometric observations with 3D MHD simulations. The spectroscopic data allows instead to study the star-disk interaction and the physics regulating the processes in the inner, gaseous disk.

    This paper is organized as follows. In Sect.~\ref{obs}, we present the observations. We update the stellar parameters of RU~Lup in Sect.~\ref{stellar_pars}. We present the analysis of the NCs and the veiling spectrum in Sect.~\ref{hotspot}, the analysis of the BC of the metallic lines in Sect.~\ref{spectroscopic_variability}, and the analysis of the TESS light curve in Sect.~\ref{TESS}. We discuss the results in Sect.~\ref{discussion} and outline our conclusions in Sect.~\ref{conclusions}.
   
    \section{Observations}
    \label{obs}
    In this section, we describe the data that we used to study RU~Lup. 
    % In the following, we define MJD$_0 \equiv 59264.336$ as reference date for the spectroscopic observations. 

    \subsection{Spectroscopy}
    The spectroscopic data were obtained with two different instruments.
    Medium-resolution optical ($4080-8900$~{\AA}) spectra were obtained with CHIRON \citep{Tokovinin2013}, an echelle spectrograph mounted on a 1.5 m telescope that is part of the Small and Moderate Aperture Research Telescope System (SMARTS) at Cerro Tololo Inter-American Observatory. 
    A total of 58 spectra were taken in three different epochs: 19 spectra in 2021, 27 spectra in 2022, and 12 spectra in 2023. Except for three observations taken at a resolving power of $R = 78000$ in 2021, the rest of the CHIRON observations have $R = 27800$. The spectra were reduced and flux-calibrated in the manner described by Walter (2018)\footnote{\url{https://www.astro.sunysb.edu/fwalter/SMARTS/CHIRON/ch_reduce.pdf}}.
    
    High-resolution ($R = 140000$) optical ($3800 - 7880$~{\AA}) spectra were obtained with the Echelle SPectrograph for Rocky Exoplanets and Stable Spectroscopic Observations \citep[ESPRESSO,][]{Pepe+2021} in the framework of the PENELLOPE program \citep{Manara+2021}.    
    Two spectra were obtained in 2021 in Pr. ID 106.20Z8.003 and five in 2022 in Pr. ID 106.20Z8.007 (PI Manara). 
    The ESPRESSO spectra were reduced by the PENELLOPE team, as is described in \citet{Manara+2021}. Telluric correction was performed using the molecfit tool \citep{Smette+2015}.

    For clarity, throughout this article each spectrum is labeled as “$\rm {ID} ~ yy.j$” where ID is either CH or ES for a CHIRON or ESPRESSO observation, respectively, $yy$ are the last two digits of the year, and $j$ is the $j^{\rm th}$ observation from that spectrograph in that year. For example, the third ESPRESSO observation from 2022 is called ES 22.3.
    The log of the spectroscopic observations is reported in Table~\ref{tab:log_specobs}.

    \subsection{Photometry}
    RU~Lup was observed with the Transiting Exoplanet Survey Satellite \citep[TESS,][]{Ricker+2014}. TESS produces short-cadence, about one-month-long light curves with a spectral response that covers the red/infrared wavelength range ($\sim 0.6-1.1 ~ \mu$m). The observation is a 200 seconds cadence light curve from Sector 65 (2023). We downloaded the light curve, extracted and reduced by the TESS Science Processing Operations Center (SPOC), from the MAST archive\footnote{\url{https://archive.stsci.edu/}}.
    The TESS Sector 65 light curve is contemporaneous to nine CHIRON observations from 2023.
    
\begin{figure*}
		\centering
		\includegraphics[width=\linewidth]{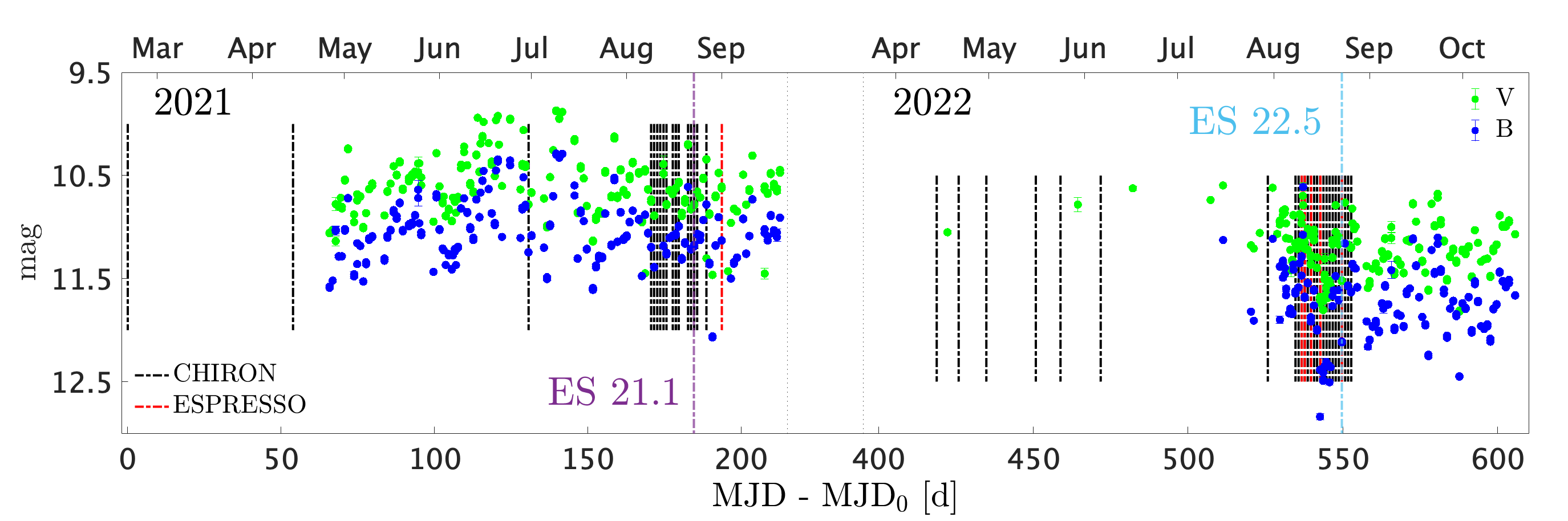}
		\caption{AAVSO $B$ (blue) and $V$ (green) photometry contemporaneous to the spectroscopic observations in 2021 and 2022. The vertical dashed lines mark the epochs of the CHIRON (black) and ESPRESSO (red) observations. The ES~21.1 and ES~22.5 spectra shown in Fig.~\ref{fig:ES21_vs_22_abs} are marked in violet and light blue. Here, MJD$_0 \equiv 59264.336$.}
		\label{fig:AAVSO_vs_spectra}
    \end{figure*}

    To supplement the spectroscopic observations with multi-band photometry, we downloaded data from the American Association of Variable Star Observers (AAVSO) International Database\footnote{\url{https://www.aavso.org/aavso-international-database-aid}}. This data set is composed of $\rm BVR_cI_c$ photometry taken in 2021 and 2022. 
    For spectroscopic data that were taken close in time to a set of photometric observations, we used the $\rm B, V, R_c$ photometry to flux-calibrate the spectra, following the procedure outlined by \citet{Armeni+2023}. We set an upper limit of 0.25 days (6 hours) on the temporal distance between the spectrum and the photometry, based on the photometric variability of the system inferred from TESS (Sect.~\ref{TESS}).
    Figure~\ref{fig:AAVSO_vs_spectra} shows the AAVSO B and V photometry, together with the spectroscopic observations from 2021 and 2022.

    In summary, we have three different epochs of spectroscopic observations: 2021, 2022, and 2023. The first two are supplemented by the AAVSO photometry, while the last one by the TESS Sector~65 light curve.
    
    \section{Stellar parameters with ESPRESSO}
    \label{stellar_pars}

    \begin{figure}
		\centering
		\includegraphics[width=\linewidth]{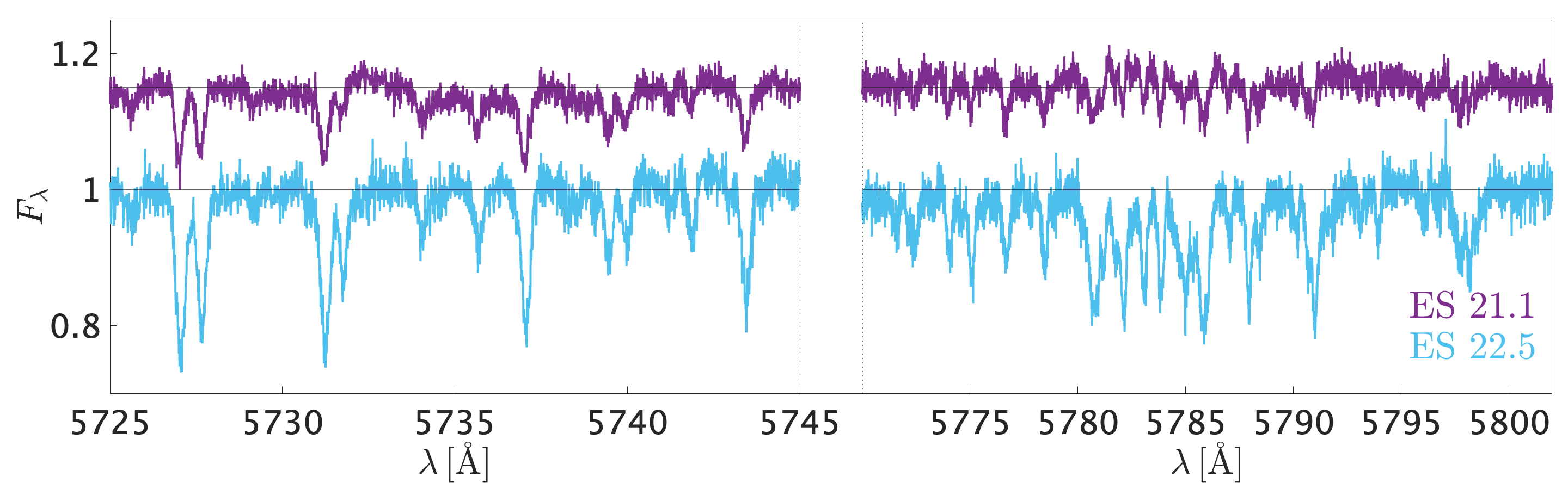}
		\caption{ES~21.1 and ES 22.5 spectra in two regions where the          photospheric absorption lines are observed. The region between         5770 and 5805 {\AA} shows the effect of line-filling emission.}
		\label{fig:ES21_vs_22_abs}
    \end{figure}
    
    The photospheric parameters of RU~Lup were previously estimated by \citet{Frasca+2017} by fitting a medium-resolution ($R \sim 17400$) X-Shooter \citep{Vernet+2011} spectrum with the ROTFIT routine \citep{Frasca+2015}. Here, we take advantage of the higher resolution of the ESPRESSO spectra to improve these measurements.
    
    Fitting a spectrum of a highly accreting CTTS such as RU~Lup is not an easy task. In some epochs, the spectrum is dominated by broad emission lines that make it difficult to identify the continuum. In addition, the photospheric lines are weaker because of veiling. An example of such an extreme veiling state is the ES~21.1 observation, shown in Fig.~\ref{fig:ES21_vs_22_abs} in violet. The region between 5770 and 5805 {\AA} shows the effect of line-filling emission of the photospheric lines.
    % , that leads to an overestimation of the continuum veiling \citep{Gahm+2008}.
    The multi-epoch monitoring of RU~Lup with ESPRESSO allowed us to observe the system in a state of lower veiling with a signal-to-noise ratio (S/N) of $\sim 40$. This observation (ES~22.5) is shown in Fig.~\ref{fig:ES21_vs_22_abs} in light blue. We focused on this spectrum to determine the stellar parameters.
    
    We applied ROTFIT to different portions of the spectrum, masking emission lines where necessary. We used as templates a library of \textit{High Accuracy Radial velocity Planet Searcher} \citep[HARPS,][]{Mayor+2003} spectra of real stars retrieved from the ESO Archive \citep[see][]{Manara+2021}. This library is mostly composed of main-sequence stars, with the exception of the early-K spectral type for which subgiant stars are also included. We avoided including weak-lined T Tauri stars because for most of them the projected rotational velocities are too high for the fitting purposes. Moreover, some of their lines are contaminated by strong chromospheric emission. Since the templates are more evolved than our targets, the values of $\log g$ determined with these templates can be overestimated. For this reason we do not report here the $\log g$ measured for RU~Lup.
    % The veiling was calculated in four 500-{\AA}-wide segments of the spectrum, from 5000 to 6500 {\AA}. % In Table~\ref{tab:stellar_accretion_pars} we report the veiling at 5500 {\AA}. 
    The stellar parameters obtained from the ROTFIT analysis of the ESPRESSO spectra are reported in Table~\ref{tab:stellar_accretion_pars}, together with the other properties of RU~Lup adopted from the literature. Thanks to the high resolution of ESPRESSO, we improved the measurement of the projected stellar rotational velocity ($v\sin i$), reducing the uncertainty by a factor of $\sim 3.5$ relative to the uncertainty given by \citet{Frasca+2017}.  
    
    Using the stellar luminosity ($L_{\star}$) obtained by \citet{Manara+2023}, we computed the stellar radius by inverting the Stefan-Boltzmann law $L_{\star} = 4\pi R_{\star}^2\sigma T_{\rm eff}^4$. The estimate of the stellar mass ($M_{\star}$) from \cite{Manara+2023}, together with $P_{\star}$ \citep{Stempels+2007} results in $R_{\rm co} = 8.26~\pm~0.65 ~ R_{\odot} = 3.64 \pm 0.88 ~ R_{\star}$. 
    %, somewhat smaller than the canonical assumption of $5~R_{\star}$ \citep[e.g.,][]{CalvetGullbring1998}.

    \begin{table}
    	\centering
    	\caption{Stellar parameters of RU\,Lup.} 
    	\begin{tabular}{ccc}
        	\hline
        	Parameter & Value & Ref. \\
        	\hline
                d & $158.9 \pm 0.7$~pc & [1] \\
                SpT & K7 & [2] \\
                $T_{\rm eff}$ & $4250 \pm 60$~K & [3] \\
                v$\sin i$ & $8.6 \pm 1.4~\rm{km~s^{-1}}$ & [3]  \\
                RV & $0.55 \pm 0.06~\rm{km~s^{-1}}$ & [3] \\                
                VF$_{5500}$ & $1.57 \pm 0.31$ & [3] \\
                $L_{\star}$ & $1.46 \pm 0.67$~$L_{\odot}$ & [4] \\
                $M_{\star}$ & $0.55 \pm 0.13$~$M_{\odot}$ & [4] \\
                $R_{\star}$ & $2.27 \pm 0.52$~$R_{\odot}$ & [3] \\
                $P_{\star}$ & $3.71 \pm 0.01$ d & [5] \\  
                $i_{\star}$ & $16 \pm 5~{}^{\rm o}$ & [3] \\ 
                $i_{\rm d}$ & $16^{+6}_{-8}~{}^{\rm o}$ & [6] \\ 
                $R_{\rm co}$ & $\sim 3.64 ~ R_{\star}$  & [3] \\             
                $A_{\rm v}$ & $\sim 0.07$~mag  & [7] \\
    	    \hline
    	\end{tabular}
    	\tablefoot{References: [1] \citet{GaiaDR3}; [2] \citet{Alcala+2017}; [3]~this work (Sect.~\ref{stellar_pars}); [4] \citet{Manara+2023}; [5] \citet{Stempels+2007}; [6] \citet{GravityColl+2021}; [7] \citet{Herczeg+2005}. }
    	\label{tab:stellar_accretion_pars} 
    \end{table}
    
    \section{Hot spot revealed by narrow component variability}
    \label{hotspot}
    The magnetospheric accretion scenario predicts the presence of hot spots on the stellar surface. The hot spots produce an excess continuum flux \citep{CalvetGullbring1998}, that can be observed photometrically \citep[e.g.,][]{Espaillat+2021}, and NCs in emission lines of species such as \ion{He}{i}, \ion{He}{ii}, \ion{Fe}{i}, \ion{Fe}{ii}, \ion{Ca}{ii}, etc. \citep{DodinLamzin2012} that can be traced spectroscopically \citep{McGinnis+2020}. Since they are formed close to the stellar surface, the NCs are expected to be rotationally modulated with $P_{\star}$. The radial velocity of the NC describes a sinusoidal curve as the star rotates
    \begin{equation}
        v_{\rm rad}(\phi) = v_0 + v \sin i \cdot \cos \theta_{\rm S} \cdot \sin \left[2\pi (\phi - \phi_{\rm S})\right]
        \label{spot_vrad}
    \end{equation}
    \citep[e.g.,][]{Sicilia-Aguilar+2015, McGinnis+2020}. Here, $v_0$ is an offset velocity that includes the stellar systemic velocity and any other velocities due to additional motions of the emitting material (such as infall), $\phi$ and $\phi_{\rm S}$ are phase angles ($0 \leq \phi, \phi_{\rm S} < 1$), and $\theta_{\rm S}$ is the latitude of the spot; that is, the angle between the position of the spot and the stellar equator. The amplitude $A = v \sin i \cdot \cos \theta_{\rm S}$ of the modulation is, thus, a fraction of $v \sin i$ and the closer the spot is to the pole, the smaller the modulation. 
    
    The study of the NC variability in EX~Lup and TW~Hya \citep{Campbell-White+2021, Sicilia-Aguilar+2023} showed how lines from different species exhibit different amplitudes and phases, and hence trace material at different positions on the stellar surface. These studies also revealed that these line-emitting regions are surprisingly stable over several years, even when the accretion rate increases \citep[e.g., during the bursts of EX~Lup,][]{Sicilia-Aguilar+2023}.
    In our case, the study of the NCs is limited by the spectral resolution of the observations. The NC of the metallic lines is typically narrower than, for instance, the NC of the \ion{He}{i} lines, making it difficult to detect in the lower resolution CHIRON spectra. Moreover, strong line blends (e.g., between the \ion{He}{i}~5016 and \ion{Fe}{ii}~5018 lines) and contamination by photospheric absorption lines (e.g., in the \ion{He}{ii}~4686 line) complicate the fit of the NC.
    For these reasons, we focused on the \ion{He}{i}~5876 line, the NC of which can always be discerned in our spectra.

    \subsection{Modulation of the \ion{He}{i}~5876 narrow component}
    \label{HeI5876_NC}
\begin{figure}
		\centering
		\includegraphics[width=\linewidth]{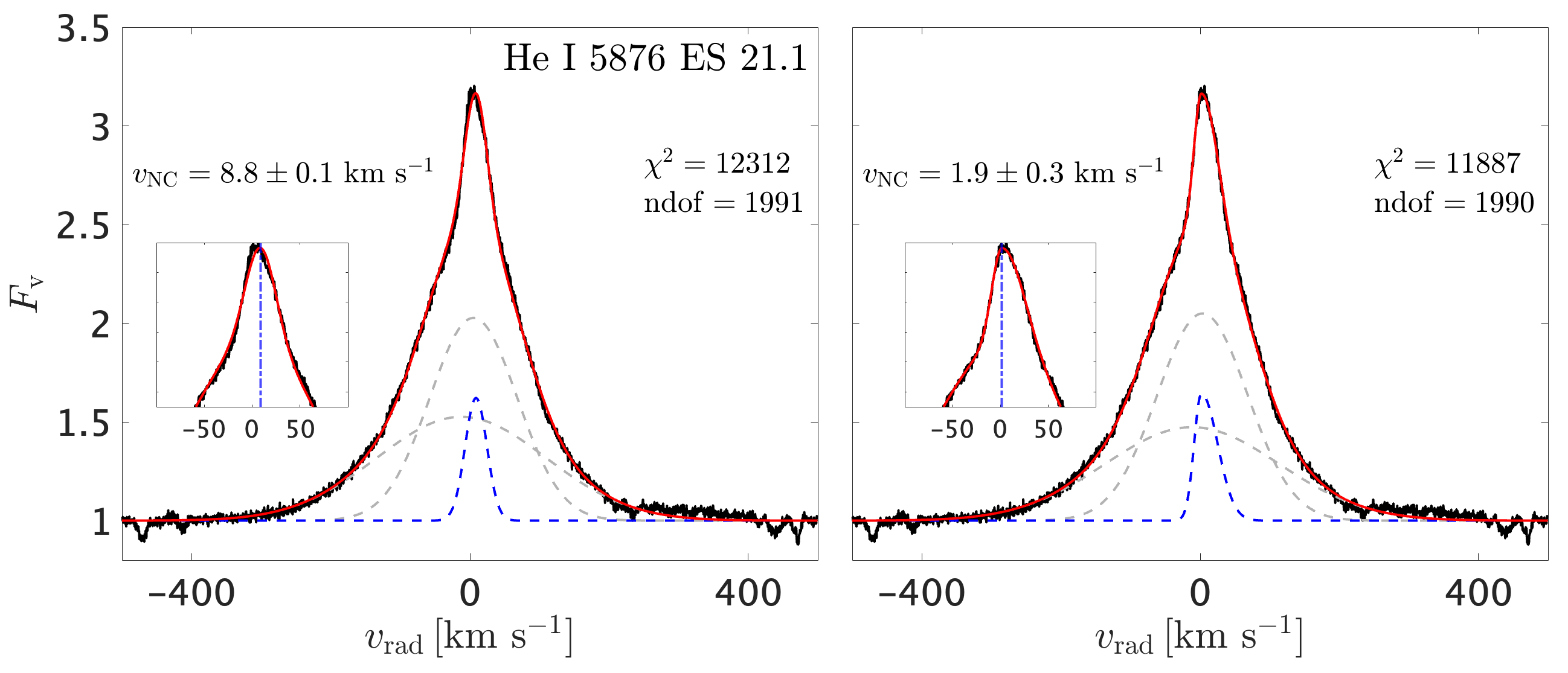}
		\caption{Best fit of the \ion{He}{i} 5876 line in ES~21.1 with two different models. Both models use three Gaussian functions. In the left panel, the Gaussian for the NC is symmetric, while in the right panel it has two different $\sigma$. See Appendix~\ref{appendix_gaussian} for details on the model. The single components are marked with dashed lines, with the blue one indicating the Gaussian for the NC. The insets are a zoom on the NC.}
		\label{fig:HeI_5876_gaussfit}
    \end{figure}

    The usual approach is to fit the line profile with two Gaussian functions, one for the BC and one for the NC \citep[e.g.,][]{Campbell-White+2021}. However, the \ion{He}{i}~5876 line profiles of RU~Lup are more complex and we had to modify the fit function (see Appendix~\ref{appendix_gaussian}). We used three Gaussian functions, two accounting for the BC and one for the NC. In addition, we allowed the Gaussian for the NC to be asymmetric, with two different widths (parameterized by the standard deviation, $\sigma$), one for the red wing ($\sigma_{\rm r}$) and the other for the blue wing ($\sigma_{\rm b}$). Figure~\ref{fig:HeI_5876_gaussfit} displays the best fit of the \ion{He}{i} 5876 line for the ES~21.1 spectrum with two different models, one using three symmetric Gaussian functions (3S) and one using the asymmetric Gaussian for the NC (2S+1A). The 2S+1A best fit demonstrates that the NC red wing is more extended than the blue wing, with $\sigma_{\rm r} = 32.9 \pm 0.5 ~ \rm{km~s^{-1}}$ and $\sigma_{\rm b} = 14.9 \pm 0.4 ~ \rm{km~s^{-1}}$. This is in agreement with formation in an infalling region. The comparison between the chi-square ($\chi^2$) of the two models indicates that the 2S+1A model better reproduces the line profile. % The radial velocity of the NC ($v_{\rm NC}$) is $6~\rm{km~s^{-1}}$ less redshifted in the asymmetric model than in the symmetric model.

\begin{figure}
		\centering
		\includegraphics[width=\linewidth]{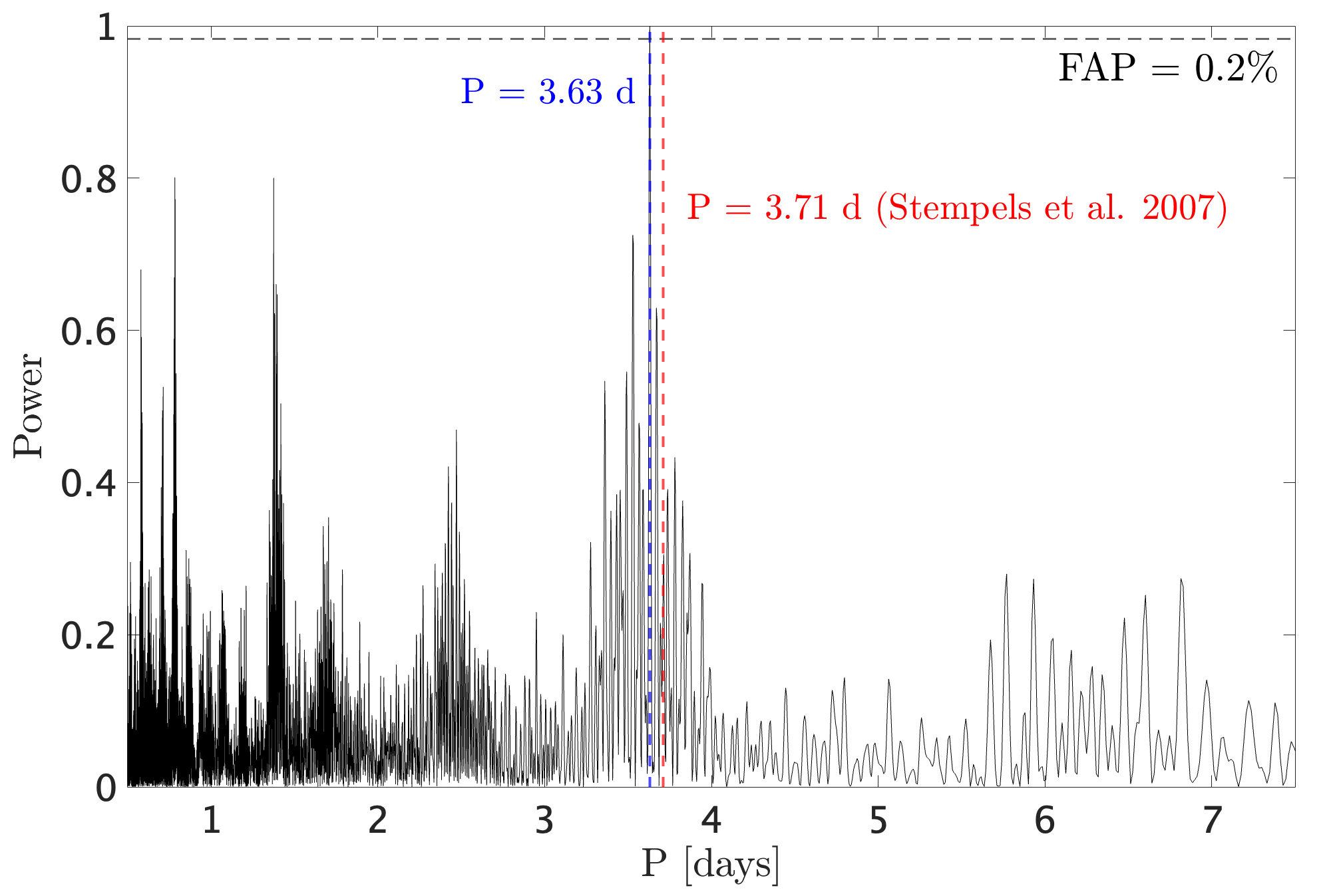}
		\caption{Lomb Scargle Periodogram of the radial velocity of the \ion{He}{i} 5876 NC.}
		\label{fig:HeI_5876_LSP}
    \end{figure}

    We fit our observations with the 2S+1A model and obtained the radial velocity of the NC ($v_{\rm NC}$), defined as in Eq.~\ref{asymm_gaussian_def}, as a function of time. Then, we computed the Lomb-Scargle Periodogram \citep[LSP,][]{Lomb1976, Scargle1982} of these radial velocity measurements. The result is shown in Fig.~\ref{fig:HeI_5876_LSP}. We detected a signal with a false alarm probabilty (FAP) of 0.2\% at a period of $3.63$ days, that differs by $\sim 2$~h from the stellar rotation period presented by \citet{Stempels+2007}. 
    The discrepancy with the \citet{Stempels+2007} period can be attributed to the intrinsic variability of the region that is traced by the NC.
    
    % Since the observation are taken on a timescale of 3 years, over which the brightness and the veiling of RU Lup varied dramatically (Sect.~\ref{veiling}), the fact that we were able to find a sinusoidal modulation indicates that the region traced by the NC, that we identify as the footprint of the magnetic field, is overall stable. 

    \begin{figure*}
		\centering
		\includegraphics[width=\linewidth]{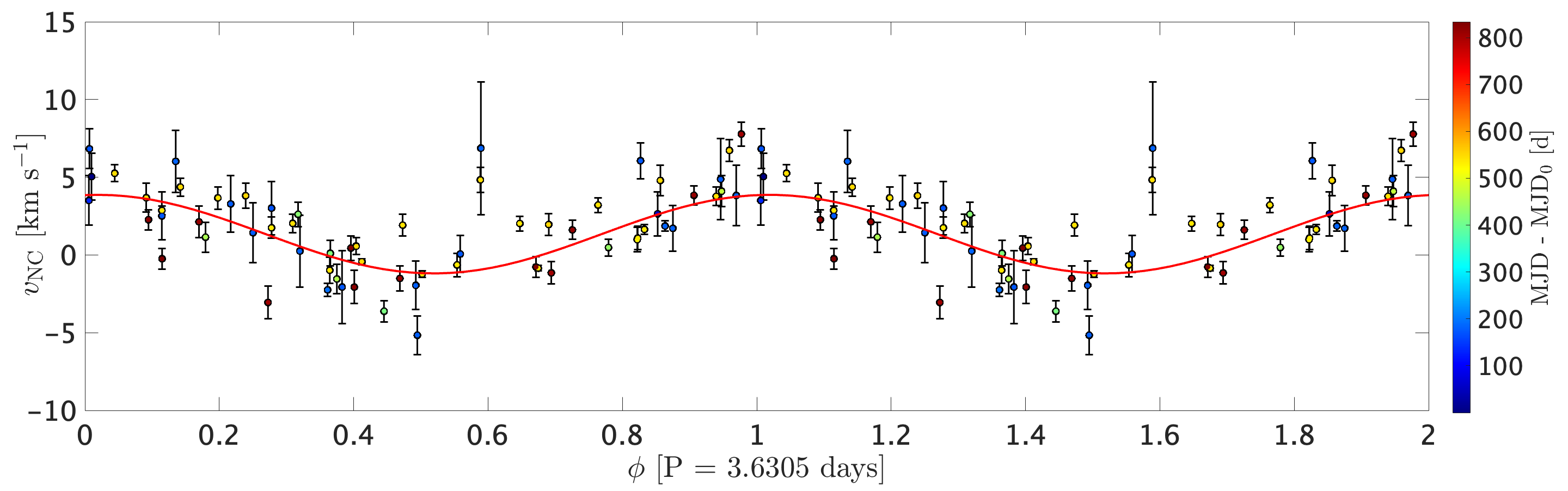}
		\caption{Phase-folded radial velocity curve of the \ion{He}{i} 5876 NC. The red line is the best fit with a sinusoidal function.}
		\label{fig:HeI_5876_phasefold}
    \end{figure*}

    Figure~\ref{fig:HeI_5876_phasefold} shows the radial velocity curve, phase-folded using the detected period and MJD$_0~\equiv ~59264.336$ as reference date for $\phi = 0$. Although there is a $\sim 1~\rm{km~s^{-1}}$ variability between different epochs and some outliers (e.g, at $\phi \approx 0.6$) the modulation is overall sinusoidal. 
    The best fit of the radial velocity curve with a sinusoidal function (Eq.~\ref{spot_vrad}) gives information about the position of the emitting region on the stellar surface \citep{Sicilia-Aguilar+2015}. We obtained $v_0 = 1.46 \pm 0.09~\rm{km~s^{-1}}$, $A = 2.53 \pm 0.13 ~\rm{km~s^{-1}}$, and $\phi_{\rm S} = 0.75 \pm 0.01$. % The amplitude is smaller than $v \sin i$, consistent with the hypothesis that the NC is formed on the stellar surface \citep{Sicilia-Aguilar+2023}. 
    Using $v \sin i = 8.6 \pm 1.4~\rm{km~s^{-1}}$ (Sect.~\ref{stellar_pars}), we derived $\theta_{\rm S} = 73 \pm 3~{}^{\rm o}$. The magnetic obliquity $\Theta = 90^{\rm o} - \theta_{\rm S}$, that is, the angle between the stellar rotation axis and the position of the NC-emitting region \citep{McGinnis+2020}, is $\sim 20^{\rm o}$, indicating that the magnetic field axis is misaligned with respect to the stellar rotation axis.
    The footprint of the magnetic field is close to the stellar pole, similar to what is observed for EX~Lup and TW~Hya \citep{Sicilia-Aguilar+2023}.

    \subsection{Narrow component of the helium and metallic lines}
    \label{NC_He_metals}
    \begin{figure}
        \centering
        \includegraphics[width=\linewidth]{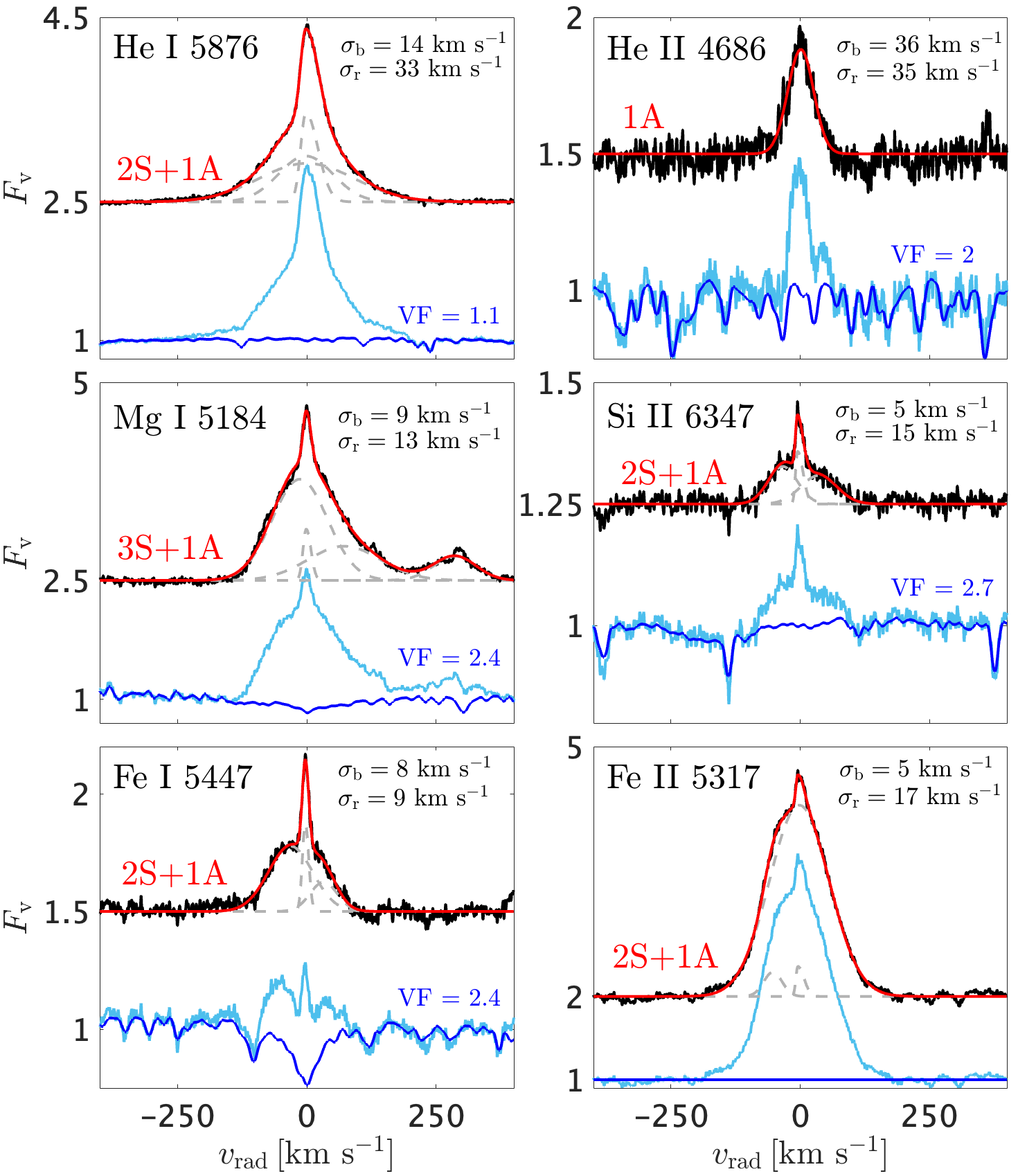}
        \caption{Selection of emission lines that have a NC. The light blue spectrum is ES~22.5. The dark blue spectra are the template that best fits the stellar spectrum (Sect.~\ref{stellar_pars}), veiled to match the depth of the photospheric lines. The photospheric subtracted spectra are shown in black, with Gaussian best fits superposed in red. The red label specifies the multiple Gaussian model used (Appendix~\ref{appendix_gaussian}).} 
        % $\sigma_{\rm b}$ and $\sigma_{\rm r}$ are referred to the NC.
        \label{fig:He_metals_NC}
    \end{figure}
    Although the resolution and the S/N of the CHIRON spectra did not allow us to study the modulation of the NC of the metallic lines, these components can be discerned in the ESPRESSO spectra with high S/N. We selected six emission lines that show a NC in ES~22.5, namely the \ion{He}{i}~5876, \ion{He}{ii}~4686, \ion{Mg}{i}~5184, \ion{Si}{ii}~6347, \ion{Fe}{i}~5447, and \ion{Fe}{ii}~5317 lines. The profile of some of these lines is severely contaminated by the underlying photospheric absorption. Therefore, we used our best fit non-accreting template to remove the stellar contribution. For each line, we adjusted the veiling to match the depth of the photospheric lines. This procedure worked for all but the \ion{Fe}{ii}~5317 line, for which the photospheric HARPS template has a gap. However, the line is strong enough that its profile can be fit even without removing the photospheric spectrum.
    
    We fit the photospheric subtracted spectra with the multiple Gaussian model introduced in Sect.~\ref{HeI5876_NC}, with the aim of deriving the parameters of the NC. For all lines, an asymmetric Gaussian was used to fit the NC.
    The results are illustrated in Fig.~\ref{fig:He_metals_NC}. The red asymmetry already observed in the \ion{He}{i}~5876 NC appears to be a common feature of the NCs of the metallic lines, with the higher-excitation lines (i.e., \ion{Si}{ii} and \ion{Fe}{ii}) being more asymmetric. Conversely, the \ion{He}{ii} line appears to be symmetric. % Since \ion{He}{ii}~4686 is the only line that does not have a BC, the asymmetry of the NC might be due to a distortion caused by the NC.
    The NCs of the helium lines are substantially broader than those of the metallic lines, suggesting that these two sets of NCs are formed in different regions of the hot spot structure. This is in agreement with the energy requirements for the formation of the helium lines. Since the upper level of the \ion{He}{ii}~4686 line has $E_{\rm j} = 51.01$~eV, this line must be formed in a region that is irradiated by the X-rays from the accretion shock. Under such conditions, magnesium and iron are expected to be ionized at least once.  

    \subsection{Anti-phase radial velocity variations}
    \label{antiphase_RV}
    
    \begin{figure}
	\centering
	\includegraphics[width=\linewidth]{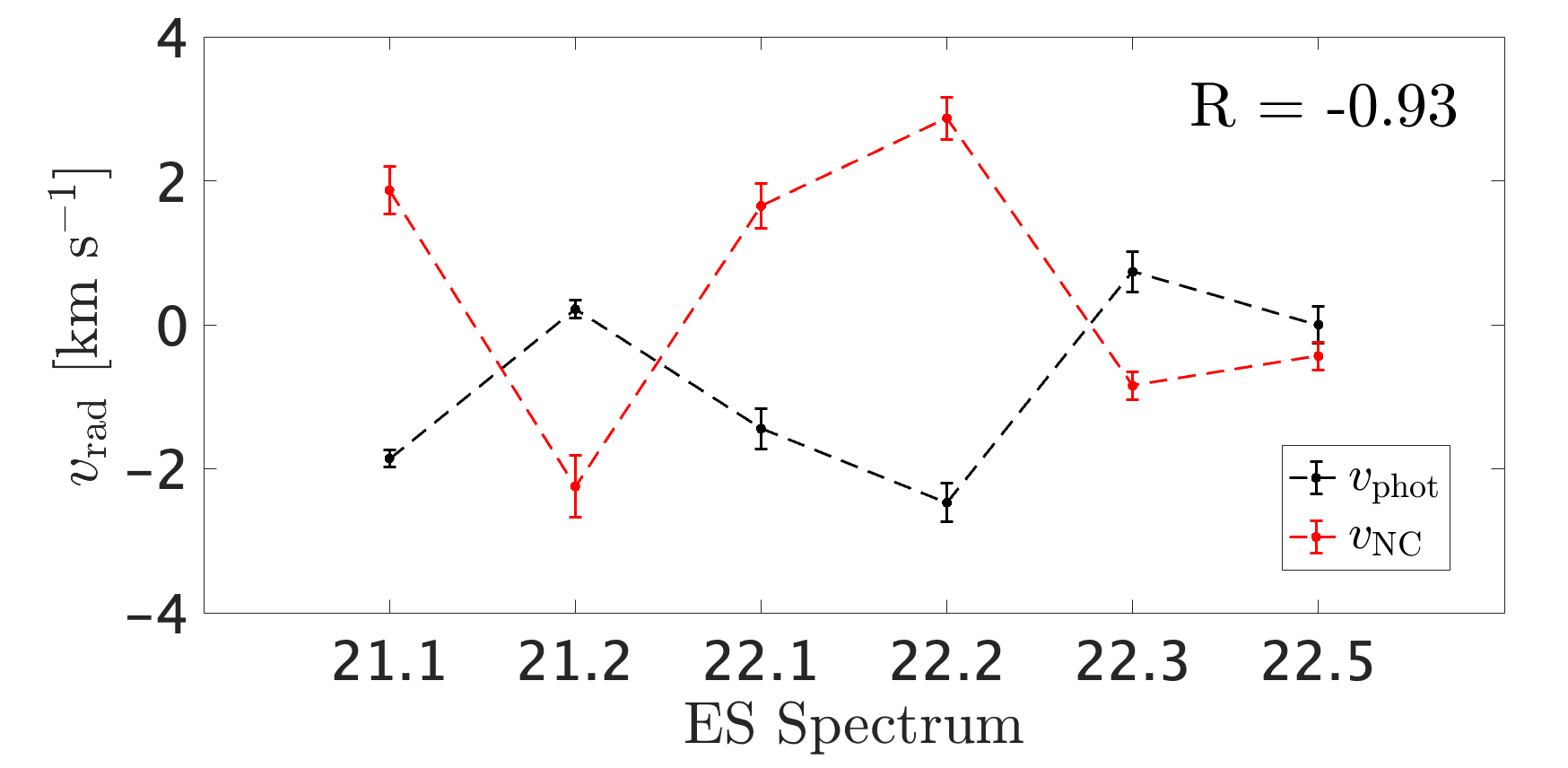}
	\caption{Anti-phase radial velocity variations of the \ion{He}{i}~5876 NC (red) and the photospheric lines (black) in the ESPRESSO spectra. R is the Pearson correlation coefficient between $v_{\rm phot}$ and $v_{\rm NC}$.}
	\label{fig:vphot_vNC_anticorr}
    \end{figure}
    We calculated the photospheric velocity relative to the ES~22.5 spectrum, $v_{\rm phot}$, by cross-correlating the latter with the other ESPRESSO spectra in the region between 5725 and 5745~{\AA}. Figure~\ref{fig:vphot_vNC_anticorr} shows how $v_{\rm phot}$ is in anti-phase with $v_{\rm NC}$. This effect was already observed for RW~Aur \citep{Petrov+2001}, DR~Tau \citep{Petrov+2011}, and EX~Lup \citep{Sicilia-Aguilar+2015}, and can be explained with the rotation of a hot spot that emits in narrow emission lines and distorts the photospheric absorption lines. This feature is offset relative to the stellar rotation axis, producing a rotational modulation of the centroids of both emission and absorption lines, which are in anti-phase \citep[e.g.,][]{Petrov+2011, Rei+2018}. Therefore, the anti-phase radial velocity variations of the \ion{He}{i}~5876 NC and the photospheric lines confirms the hypothesis that the NCs are produced in a hot spot on the stellar surface in RU~Lup.

    \subsection{Veiling spectrum: Continuum and line emission}
    \label{veiling}
    
    \begin{figure*}
	\centering
	\includegraphics[width=\linewidth]{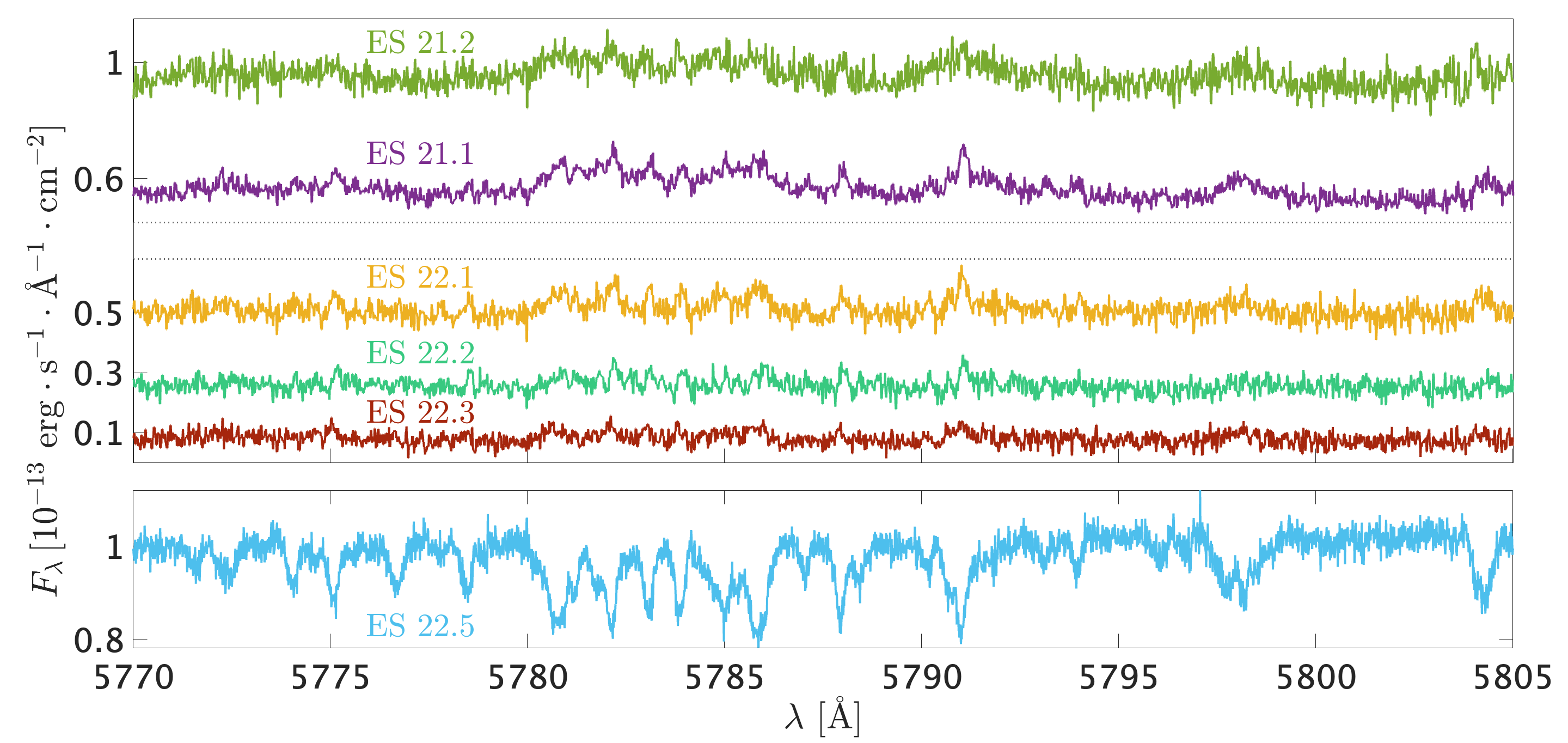}
	\caption{Flux-subtraction of the ES~22.5 spectrum from the other ESPRESSO spectra. The top panel shows the five subtracted spectra median filtered to three points. ES~22.5 is shown in the bottom panel for reference.}
	\label{fig:veiling_disentanglement}
    \end{figure*}
    We focused on the high-resolution ESPRESSO spectra to study the veiling spectrum. Figure~\ref{fig:ES21_vs_22_abs} shows how the photospheric lines are not only veiled by an excess continuum, but also by narrow line-filling emission likely coming from the hot spot (Sect.~\ref{antiphase_RV}). 
    
    Since the ESPRESSO spectra are absolutely flux-calibrated, it is possible to disentangle the continuum and the line emission contribution to the veiling. To do this, we subtracted ES~22.5 from the other ESPRESSO observations, after correcting for shifts in radial velocity by cross-correlating the spectra. This procedure is a flux-calibrated version of the one employed by \citet{Herczeg+2023} to measure the veiling in TW~Hya.
    The result is illustrated in Fig.~\ref{fig:veiling_disentanglement} for the region around 5800~{\AA}.
    
    Under the assumption that the ES~22.5 observation represents the photospheric spectrum plus a featureless continuum with a VF of VF$_{\rm T} = 1.57$ (Sect.~\ref{stellar_pars}), the subtraction removes the photospheric spectrum and reveals the veiling spectrum, which consists of two components: continuum and line emission. The continuum veiling fraction relative to ES~22.5 (VF$_{\rm C}$) can be directly calculated as the ratio between the average fluxes of the subtracted spectrum and the ES~22.5 spectrum in the region between 5800 and 5803~{\AA}, where the spectra are free of absorption lines. This means that in the subtracted spectra, that region represents the excess continuum relative to ES~22.5.

    In Appendix~\ref{appendix_veiling}, we calculated the relative veiling, VF$_{\rm rel}$, between any given spectrum and the template ES~22.5 by finding the value that best fits the normalized spectrum assuming that the photospheric lines are weakened only by an excess continuum.
    This value is much higher than the value of VF$_{\rm c}$ that we find above for the same wavelength region. As an example, we calculated VF$_{\rm rel} = 2.91 \pm 0.11$ and VF$_{\rm C} = 0.52 \pm 0.03$ for the ES~21.1 observation. The values of VF$_{\rm C}$ and VF$_{\rm rel}$ are different because of the effect of line-filling emission, which reduces the depth of the photospheric lines in normalized spectra in the same way as a diluting continuum (see, e.g., Fig.~\ref{fig:veiling_ES21.1}). This result highlights the importance of the contribution of the line-filling emission to the veiling in highly veiled CTTSs, as already discussed by \citet{Gahm+2008}, \citet{Petrov+2011}, and \citet{Rei+2018}.

    \section{Spectroscopic variability of the metallic lines}
    \label{spectroscopic_variability}
    \begin{figure*}
	\centering
	\includegraphics[width=\linewidth]{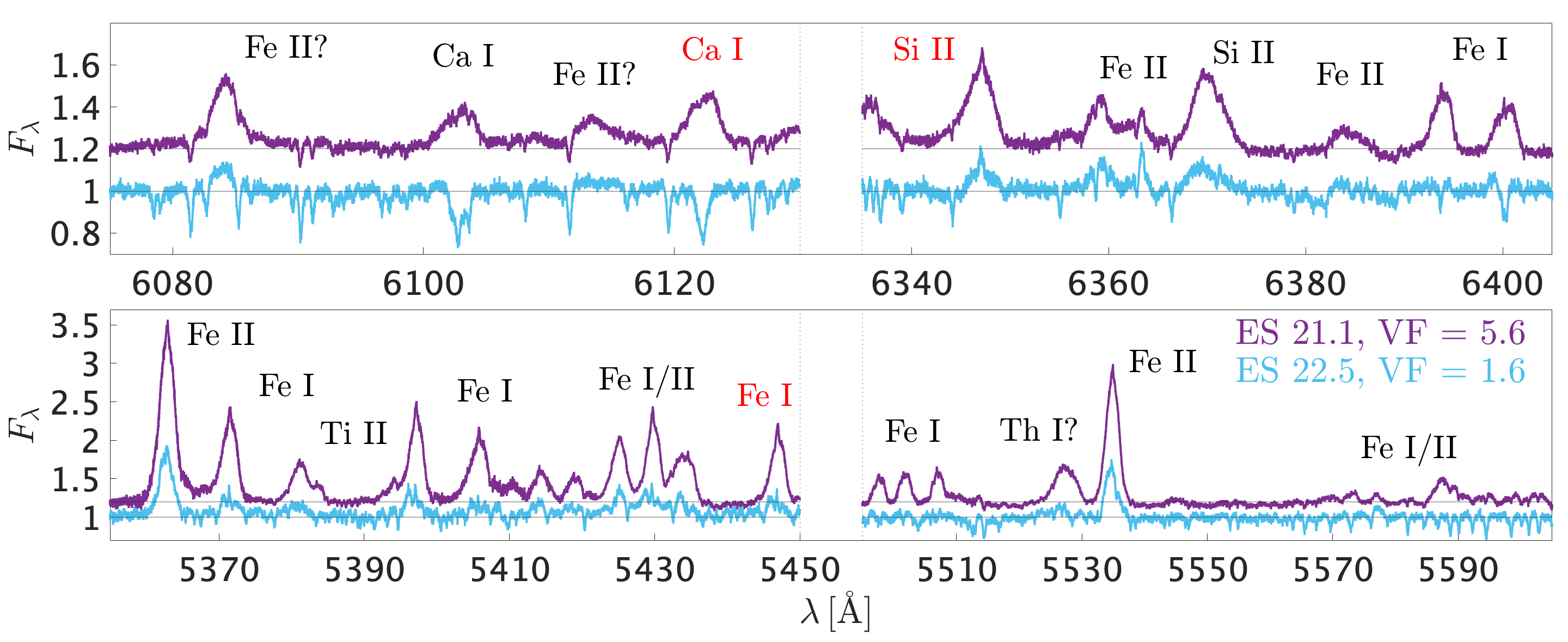}
	\caption{Comparison between the ES~21.1 and ES 22.5 spectra in regions dominated by metallic emission lines. Species producing line emission are marked. The \ion{Fe}{i} 5447, \ion{Si}{ii} 6347, and \ion{Ca}{i} 6122 lines used in Sect.~\ref{CHIRON_TESS} are marked in red.}
	\label{fig:ES21_vs_22_emission}
    \end{figure*}
    The two spectra in Fig.~\ref{fig:ES21_vs_22_abs} represent two different states of veiling of RU~Lup. We estimated the veiling of the ES~21.1 observation, as is explained in Appendix~\ref{appendix_veiling}, obtaining $\rm{VF} = 5.6 \pm 0.9$. For the ES~22.5 spectrum, we derived $\rm{VF} = 1.6 \pm 0.3$ (Sect.~\ref{stellar_pars}). Figure~\ref{fig:ES21_vs_22_emission} shows the differences in the emission line spectrum between these two observations. Emission lines from \ion{Fe}{ii} (e.g, \ion{Fe}{ii}~5363 and 5535) are present in both spectra, but they are stronger relative to the continuum in ES~21.1. The emission line spectrum is richer in ES~21.1, with a large number of transitions, mostly from neutral species such as \ion{Fe}{i} and \ion{Ti}{i}, appearing in emission. 
    These metallic lines have a BC with a full width at half maximum (FWHM) of $\sim 150 ~ \rm{km~s^{-1}}$. In Fig.~\ref{fig:ES21_vs_22_emission} we marked in red the \ion{Fe}{i} 5447, \ion{Si}{ii} 6347, and \ion{Ca}{i} 6122 lines, which are not blended with other lines and are clear examples of how the line strength increases relative to the continuum in the spectrum with higher veiling. The \ion{Ca}{i} doublet ($\lambda\lambda~6103, 6122$) has the most striking variability between the two observations, being strongly in absorption in ES~22.5 but completely filled in with emission in ES~21.1. The comparison between the ES~21.1 and 22.5 spectra indicates that the strength of these lines can be used as proxy of states of the accretion state.

    Several studies \citep[e.g.,][]{Petrov+2001, Sicilia-Aguilar+2012, Sicilia-Aguilar+2023} showed that the BC of the metallic lines is formed in the circumstellar environment of the star.    
    % The correlation between the EW of the BMLs and $\dot{M}_{\rm acc}$, derived from the \ion{He}{I} luminosity, indicates that these emission lines are linked to the accretion process. 
    However, these lines must be formed in a different region with respect to, for instance, the \ion{He}{i}~lines. Iron and calcium are expected to be ionized in the high temperature conditions required for the formation of the \ion{He}{i}~lines \citep{Armeni+2023}. 

    \begin{figure}
		\centering
		\includegraphics[width=\linewidth]{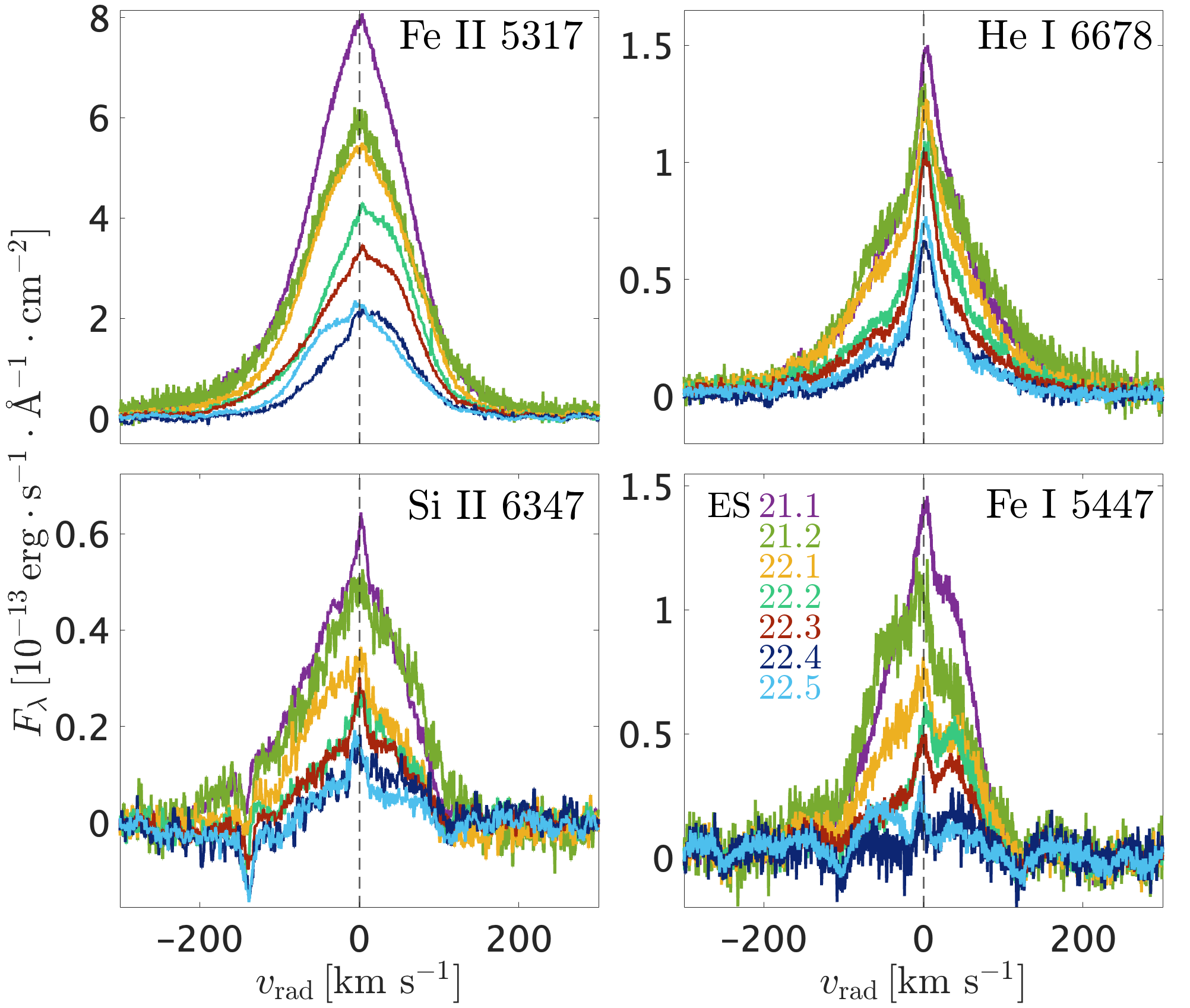}
		\caption{Continuum subtracted profiles of the \ion{Fe}{ii}~5317, \ion{He}{i}~6678, \ion{Si}{ii}~6347, and \ion{Fe}{i}~5447 lines in the ESPRESSO spectra.}
		\label{fig:FeII_HeI_SiII_FeI_ES}
    \end{figure}
    
    To understand the difference between the region of formation of these species, we selected four emission lines to analyze their BC. These lines are the \ion{He}{i}~6678, \ion{Fe}{i}~5447, \ion{Fe}{ii}~5317, and \ion{Si}{ii}~6347 lines. The $\lambda 5447$ and $\lambda 5317$ lines are among the few iron lines that are isolated and not blended with other emission lines. 
    We chose the $\lambda 6678$ line instead of the $\lambda 5876$ line for \ion{He}{i} because it is less optically thick, hence a better tracer of the gas dynamics. The \ion{Si}{ii}~6347 line was selected because its profile is not contaminated by photospheric absorption, given the high energy of its lower level ($E_{\rm i} = 8.12$~eV). The lines are plotted in absolute flux in Fig.~\ref{fig:FeII_HeI_SiII_FeI_ES}. 
    % All the transitions are stronger in 2021. 
    % However, while the \ion{He}{i} line has roughly the same luminosity in ES~21.1 and 21.2, the \ion{Fe}{ii}~5317 line is much stronger in ES~21.1, just like VF$_{\rm L}$ (Sect~\ref{veiling}). 
    % In the three spectra where it is strongest, that is, ES~21.1, 21.2, and 22.1, the \ion{Fe}{ii} line has a profile with a triangular core and broad wings extending up to $\pm 250~\rm{km~s^{-1}}$. In this situation, information about the gas dynamics from this line is lost, probably because the line becomes too optically thick. This is evident when comparing the \ion{Fe}{ii} and \ion{Si}{ii} lines. In ES~22.1, \ion{Si}{ii}~6347 is skewed to the red, whereas this asymmetry is not visible in \ion{Fe}{ii}. 
    % However, when the accretion rate decreases, velocity shifts can be better observed in \ion{Fe}{ii} than in \ion{Si}{ii} due to its higher S/N. The BC of the \ion{Fe}{ii}~5317 line is skewed to the blue with a redshifted line center in ES~22.2, 22.3, and 22.4. In ES 22.5 the situation is reversed. 
    
    \begin{figure}
        \begin{center}
            \parbox{9cm}
            {
            {\includegraphics[width=0.5\textwidth]{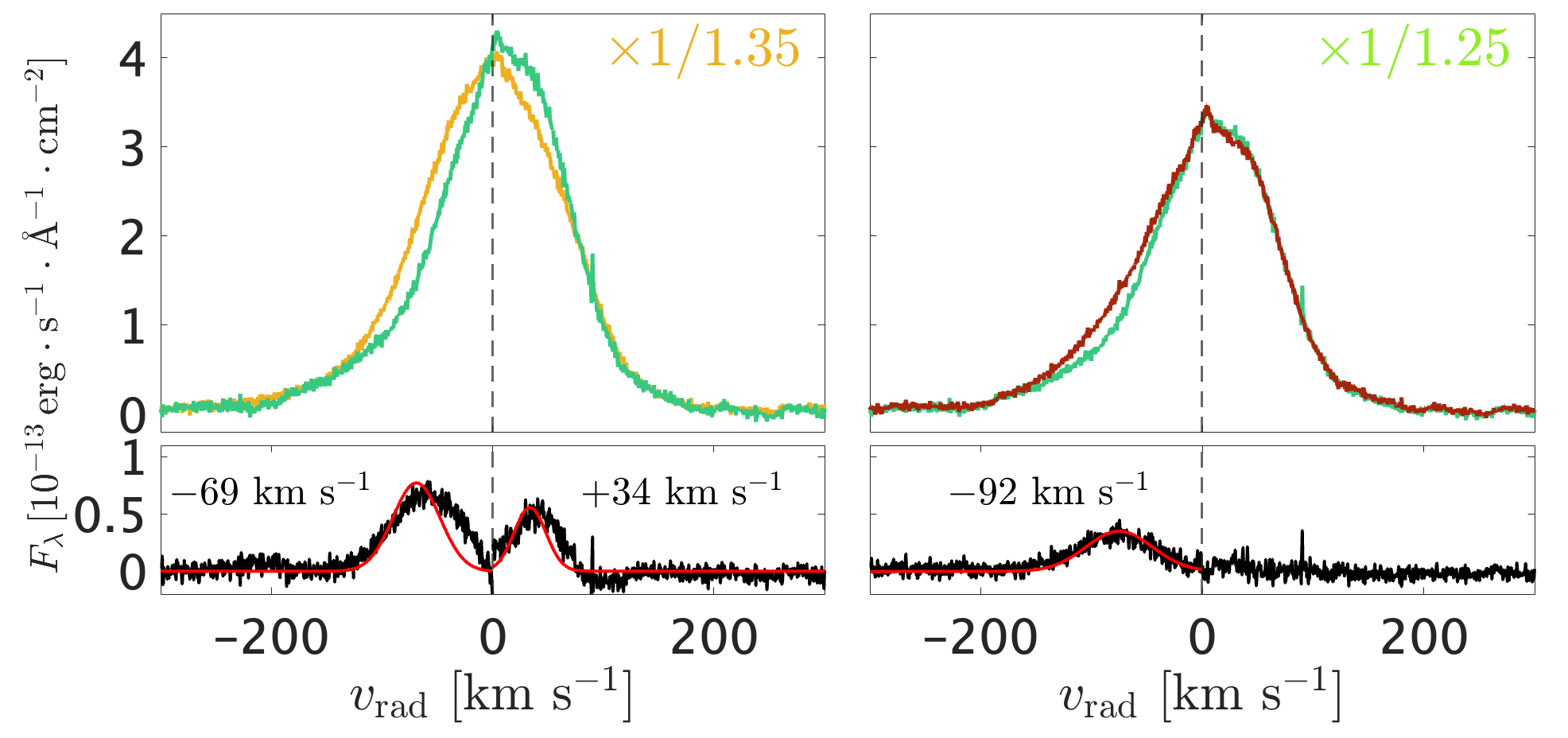}}        
            }
            \parbox{9cm}
            {
            \parbox{4.8cm}{\includegraphics[width=0.27\textwidth]{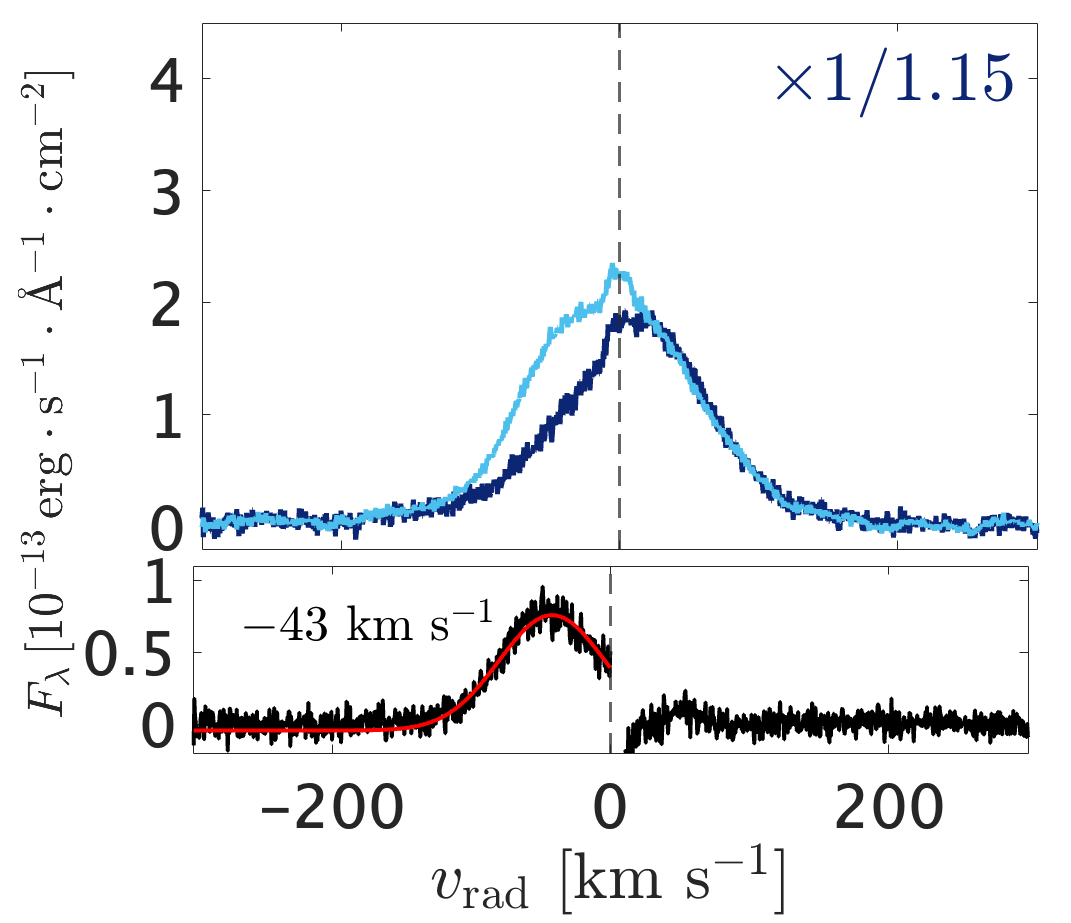}}
            \hspace{0cm}
            \parbox{4cm}{\caption{Subtraction between consecutive \ion{Fe}{ii}~5317 ESPRESSO spectra. The color code is the same as in Fig.~\ref{fig:FeII_HeI_SiII_FeI_ES}. The profiles were rescaled by a constant (upper right corner). The bottom panels show the subtracted profiles, with Gaussian fits superposed.}
	\label{fig:FeII_ES22_subtraction}}
            }
        \end{center}
    \end{figure}

    The comparison between the \ion{Fe}{ii}~5317 line profiles in the ESPRESSO spectra shows the existence of a component with a variable radial velocity. This is illustrated in Fig.~\ref{fig:FeII_ES22_subtraction}, where three sets of two consecutive ESPRESSO observations are subtracted from each other in the \ion{Fe}{ii}~5317 line. Before subtracting, we rescaled the line profiles so that the wings matched. % The ES~22.1 and ES~22.2 spectra were acquired only $\sim 1$~day apart in time. 
    The subtraction reveals that the component has velocity centroids ranging between $\sim - 90$ and $\sim + 35~\rm{km~s^{-1}}$ and it is responsible for the variations in the shape of the line.

    % The subtraction reveals the motion of a feature that is blueshifted by $\sim 70~\rm{km~s^{-1}}$ in ES~22.1 and redshifted by $\sim 35~\rm{km~s^{-1}}$ in ES~22.2. After $\sim 2$~days, a blueshifted component centered at $\sim -90~\rm{km~s^{-1}}$ appears in ES~22.3, while the red wing is stable. The comparison between ES~22.4 and ES~22.5 (for which the time difference is $\sim 7$~d) highlights a feature with a line center at $-43 ~\rm{km~s^{-1}}$ that shows up in ES~22.5.

    \begin{figure}
        \begin{center}
            \parbox{9cm}
            {
            \hspace{-0.125cm}
            {\includegraphics[width=0.5\textwidth]{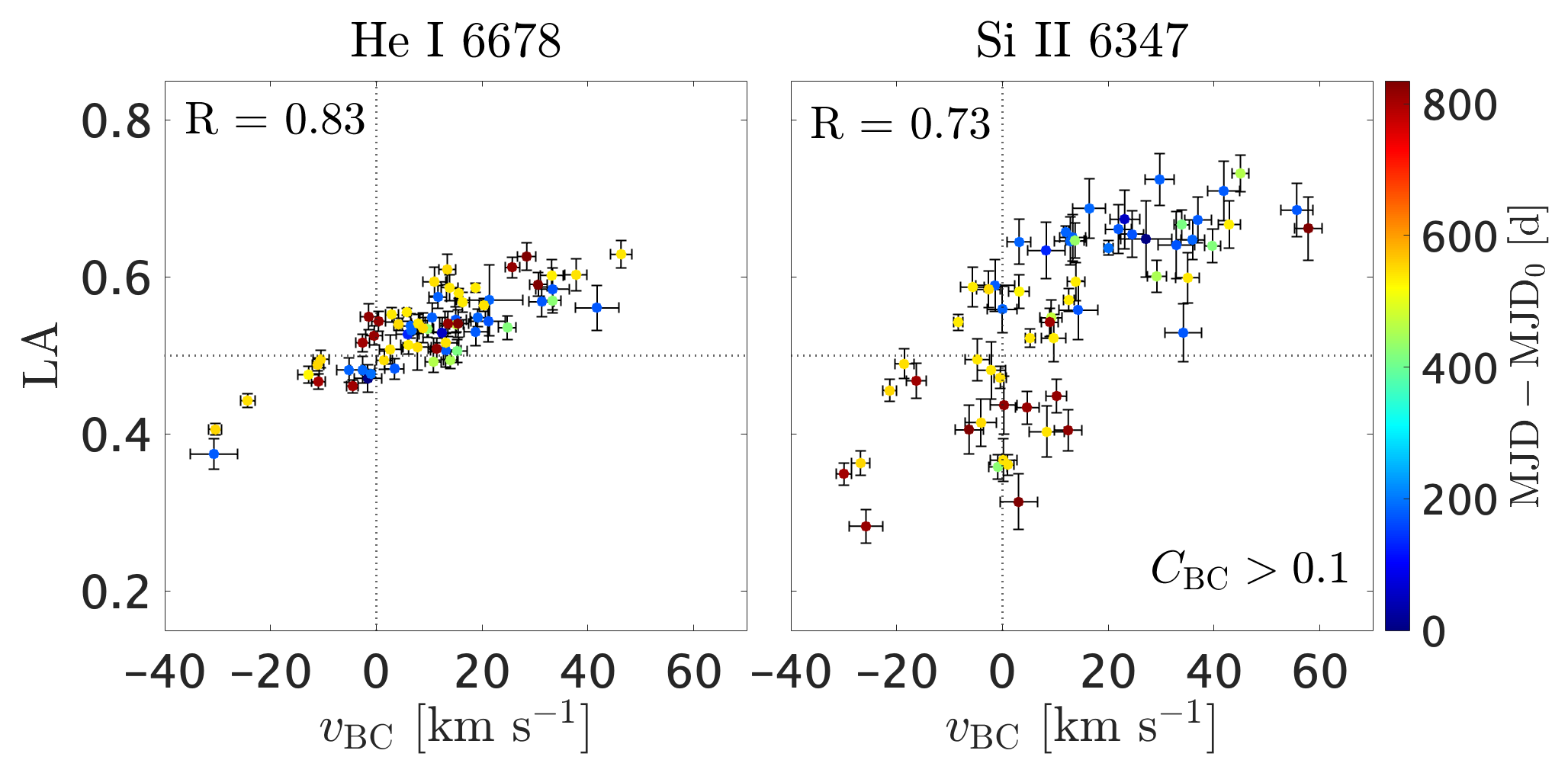}}        
            }
            \parbox{9cm}
            {
            \parbox{4.5cm}{\includegraphics[width=0.25\textwidth]{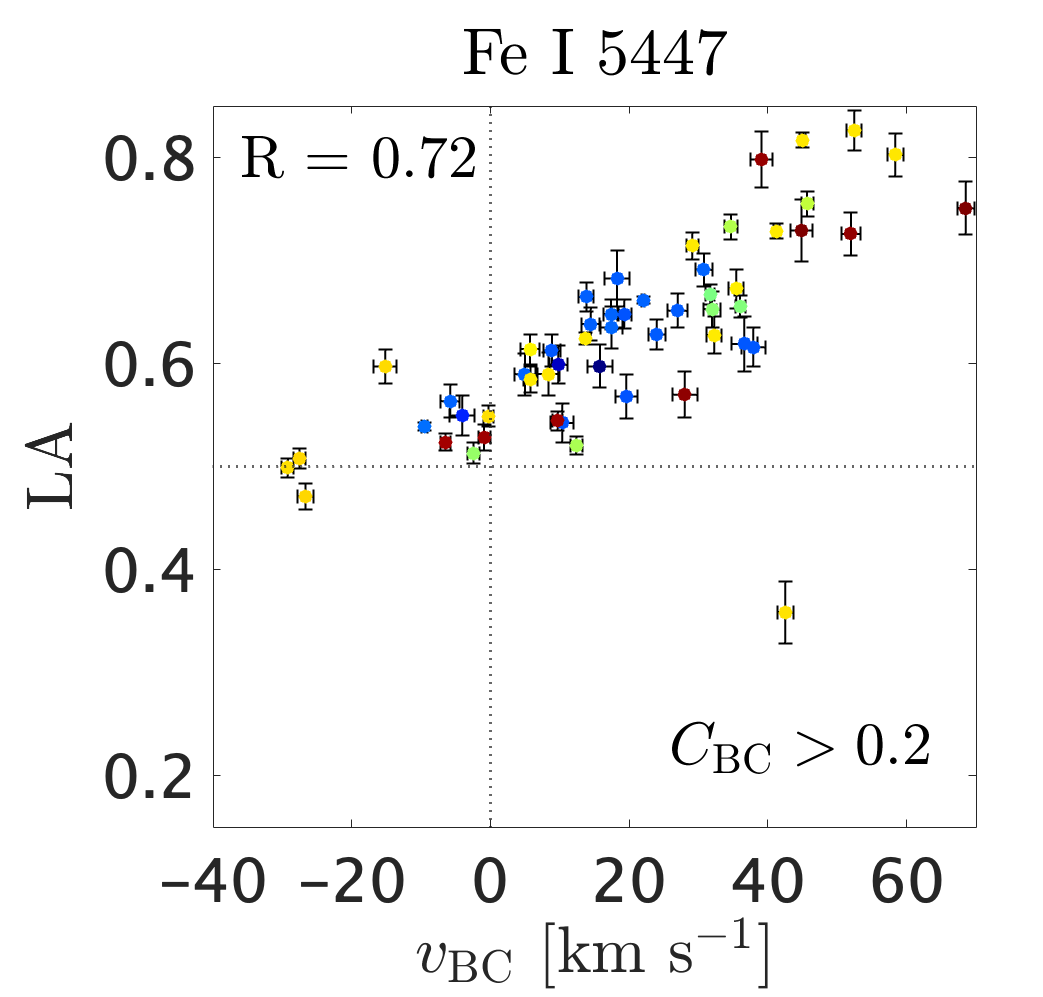}}
            \parbox{4cm}{\caption{Relation between $v_{\rm BC}$ and LA for the \ion{He}{i}~6678, \ion{Si}{ii}~6347, and \ion{Fe}{i}~5447 lines. R is the Pearson correlation coefficient.} \label{fig:BMLs_vBC_vs_LA}}
            }
        \end{center}
    \end{figure}
    
    The variability of the asymmetry of the BC suggests the presence of a non-axisymmetric structure rotating around the star. To further explore this scenario, we fit the other three selected lines with Gaussian functions. For all three lines, we used an asymmetric Gaussian to fit the BC. The NC of the \ion{He}{i}~6678 line can be always discerned in our observations. Therefore, we fit the \ion{He}{i} line profiles with a 1S+1A model. Conversely, the NC of the \ion{Fe}{i} and \ion{Si}{ii} lines is sometimes absent or it cannot be discerned in observations with lower S/N. For this reason, we masked the profiles of these lines between -15 and +15 $\rm{km~s^{-1}}$ and fit the remaining line profile with a 1A model. The fit parameters for the BCs are the strength of the component relative to the continuum ($C_{\rm BC}$), its velocity ($v_{\rm BC}$), and the widths of the blue and red wings ($\sigma_{\rm b}$ and $\sigma_{\rm r}$). From the widths, we derived the FWHM and the line asymmetry to the blue (LA) as $\rm{FWHM} = \sqrt{2 \ln 2} \cdot (\sigma_{\rm b} + \sigma_{\rm r})$ and $\rm{LA} = \sigma_{\rm b}/(\sigma_{\rm b} + \sigma_{\rm r})$. 
    When the line was weak relative to the continuum, the fit did not converge. For this reason, we excluded fits with $C_{\rm BC} < 0.2$ for \ion{Fe}{i}~5447 and $C_{\rm BC} < 0.1$ for \ion{Si}{ii}~6347.
    The results of the best fits are displayed in Fig.~\ref{fig:BMLs_vBC_vs_LA} in the form of the relation between $v_{\rm BC}$ and LA for each emission line. There is a high (Pearson R $> 0.70$) positive correlation between these two parameters for each line, that is, the higher LA, the more the line center is redshifted.
    The \ion{He}{i}~6678 BC is less influenced by line shifts, it is most of the times asymmetric to the blue (LA $> 0.5$) and redshifted. Also the \ion{Fe}{i}~5547 BC is redshifted in 42 out of 53 observations, and it is sometimes highly blue-skewed (with LA $\gtrsim 0.7$). The \ion{Si}{ii}~6347 BC is the only one with significant asymmetry to the red (LA $< 0.5$). All the lines have $v_{\rm BC}$ between $-30$ and $+60 ~ \rm{km~s^{-1}}$.

    % \begin{figure}
    %		\centering
    %	    \includegraphics[width=\linewidth]{figs/BMLs_LA_corr}
    %		\caption{LA of the \ion{Si}{ii}~6347 line vs. LA of the \ion{Fe}{i}~5447 and \ion{He}{i}~6678 lines.}
    % 		\label{fig:BMLs_LA_corr}
    % \end{figure} 
    % Figure~\ref{fig:BMLs_LA_corr} shows the relation between the LA of the \ion{Si}{ii}~6347 line and the other two lines. Although there is low correlation ($|R| < 0.2$) in both cases, the \ion{Si}{ii} vs. \ion{He}{i} plot has less scatter than the \ion{Si}{ii} vs. \ion{Fe}{i} plot. 

    \begin{figure}
        \centering
        \includegraphics[width=\linewidth]{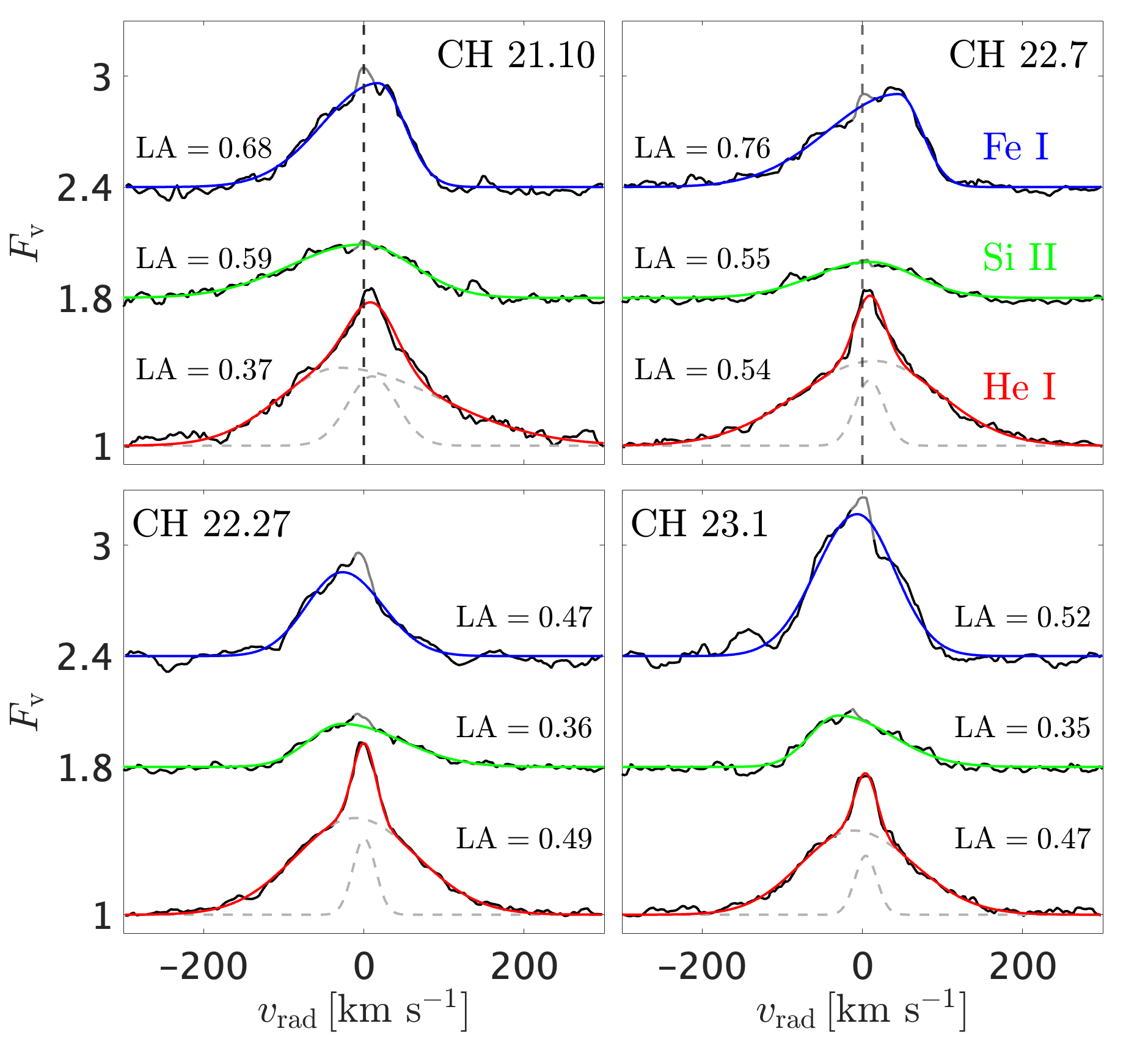}
        \caption{Selected observations showing the variability of the asymmetry of the \ion{He}{i}~6678, \ion{Si}{ii}~6347, and \ion{Fe}{i}~5447 lines. Here, LA is the line asymmetry to the blue.}
        \label{fig:BMLs_examples}
    \end{figure} 
    A selection of four observations that highlight the variability of the line asymmetry is shown in Fig.~\ref{fig:BMLs_examples}. 
    % In CH~21.10 and CH~22.7 the \ion{Fe}{i} line is more blue-skewed than the \ion{Si}{ii} and \ion{He}{i} lines. In CH~21.10, the \ion{He}{i} BC appears to be even skewed to the red, although the best fit of the BC might be influenced by the NC, that is broader in this observation. On the other hand, in the CH~22.7 and CH~23.1 observations the \ion{Fe}{i} and \ion{He}{i} BCs are symmetric ($\rm{LA} \approx 0.5$) but the \ion{Si}{ii} is strongly asymmetric to the red ($\rm{LA} \approx 0.35$).
    % The \ion{He}{i}~6678 BC is symmetric, except in CH~21.10 where it is skewed to the red. 
    In CH~21.10 and CH~22.7 the \ion{Si}{ii} BC is blue-skewed, while in CH~22.27 and CH~23.1 it is red-skewed. 
    The change from red to blue asymmetry is in agreement with what is observed in the \ion{Fe}{ii}~5317 line and further supports the hypothesis of a non-axysimmetric flow rotating around the star.
    Different transitions have different line asymmetries in the same observation, suggesting that although these emission lines are produced in the accretion flow, they trace different regions of it. This is confirmed by the different FWHM of the lines, which is on average $\sim 250$, $190$, and $173 ~ \rm{km~s^{-1}}$, for the \ion{He}{i}, \ion{Si}{ii}, and \ion{Fe}{i} lines, respectively. 
    These values are compatible with a formation in a stratified flow, as observed for HM~Lup \citep{Armeni+2023}. The \ion{He}{i}~6678 line has the most extreme excitation conditions among the three lines, since its upper level has $E = 23.07$~eV. Therefore, it must be formed closer to the star, where the gas is likely irradiated by the X-ray radiation from the shock and has higher velocity, than the \ion{Si}{ii} and \ion{Fe}{i} lines. 
    % In order to interpret the line variability and the different velocities observed in the BCs, we analyzed the TESS light curves looking for rotational timescales.
    
    \section{TESS light curve}
    \label{TESS}

    \begin{figure*}
	\centering
	\includegraphics[width=\linewidth]{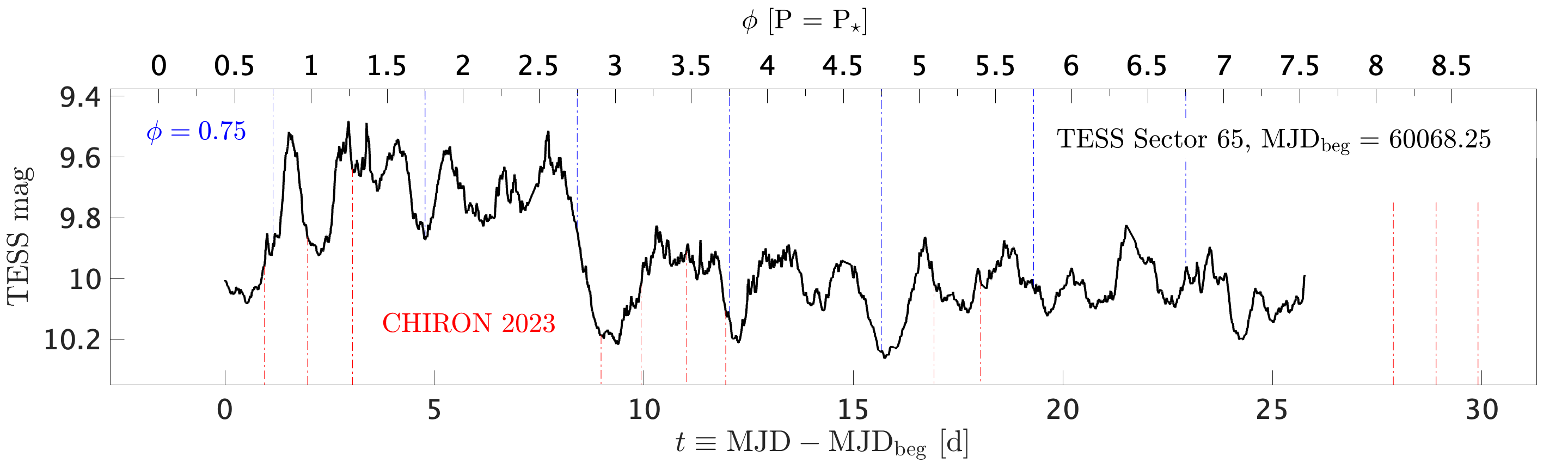}
	\caption{TESS light curve of RU~Lup from Sector~65 (2023). The light curve is plotted as a function of the time $t$ from the beginning of the observation (MJD$_{\rm beg} \equiv 60068.25$). The top axis show the phase $\phi$ computed with $P = 3.63$~days (i.e., the spot period, Sect.~\ref{HeI5876_NC}). The reference date for $\phi = 0$ is MJD$_0~\equiv ~59264.336$. The blue lines mark the phase $\phi_{\rm S}$ (from Sect.~\ref{HeI5876_NC}), where the maximum contribution from the hot spot is expected. The red lines mark the epochs of the CHIRON spectroscopic observations.}
	\label{fig:TESS_2019_2023}
    \end{figure*}
    
    The TESS Sector~65 light curve of RU~Lup is displayed in Fig.~\ref{fig:TESS_2019_2023}. We converted the TESS pre-search data conditioning simple aperture photometry (PDCSAP) flux ($F$) into TESS magnitudes $T$ using the relation $T ~ = ~-2.5\log_{10}F~+~\rm{ZP}$, where $\rm{ZP} = 20.44$ is the TESS Zero Point magnitude \citep{Vanderspek+2018}. 
    The time $t$ is computed relative to the MJD of the beginning of the observation (MJD$_{\rm beg} = 60068.25$).
    
    If the hot spot has a photometric signature, we expect its maximum contribution at phase $\phi_{\rm S}$ (Sect.~\ref{hotspot}), marked with dashed blue lines in Fig.~\ref{fig:TESS_2019_2023}. This phase is a quarter of period after the maximum blueshift and a quarter of period before the maximum redshift of the hot spot, hence it is the position where the spot area projected along the line of sight is maximum.
    % We used $\phi_{\rm S} = 0.75$ (from the global best fit) for the 2019 TESS light curve and $\phi_{\rm S} = 0.70$ (from the best fit of the 2023 spectra) for the 2023 TESS light curve.
    For most cycles this phase does not coincide with a maximum in the TESS light curve. 
    % The Sector~65 observation is contemporaneous to the CHIRON spectra from 2023, that are part of the data set used in Sect.~\ref{hotspot} to find the periodic modulation of $v_{\rm NC}$. 
    This suggests that either the hot spot does not contribute significantly to the photometry or the period detected from the analysis of the NC is not the actual rotation period of the spot. We further discuss this issue in Sect.~\ref{discussion}.

    \begin{figure*}
	\centering
	\includegraphics[width=\linewidth]{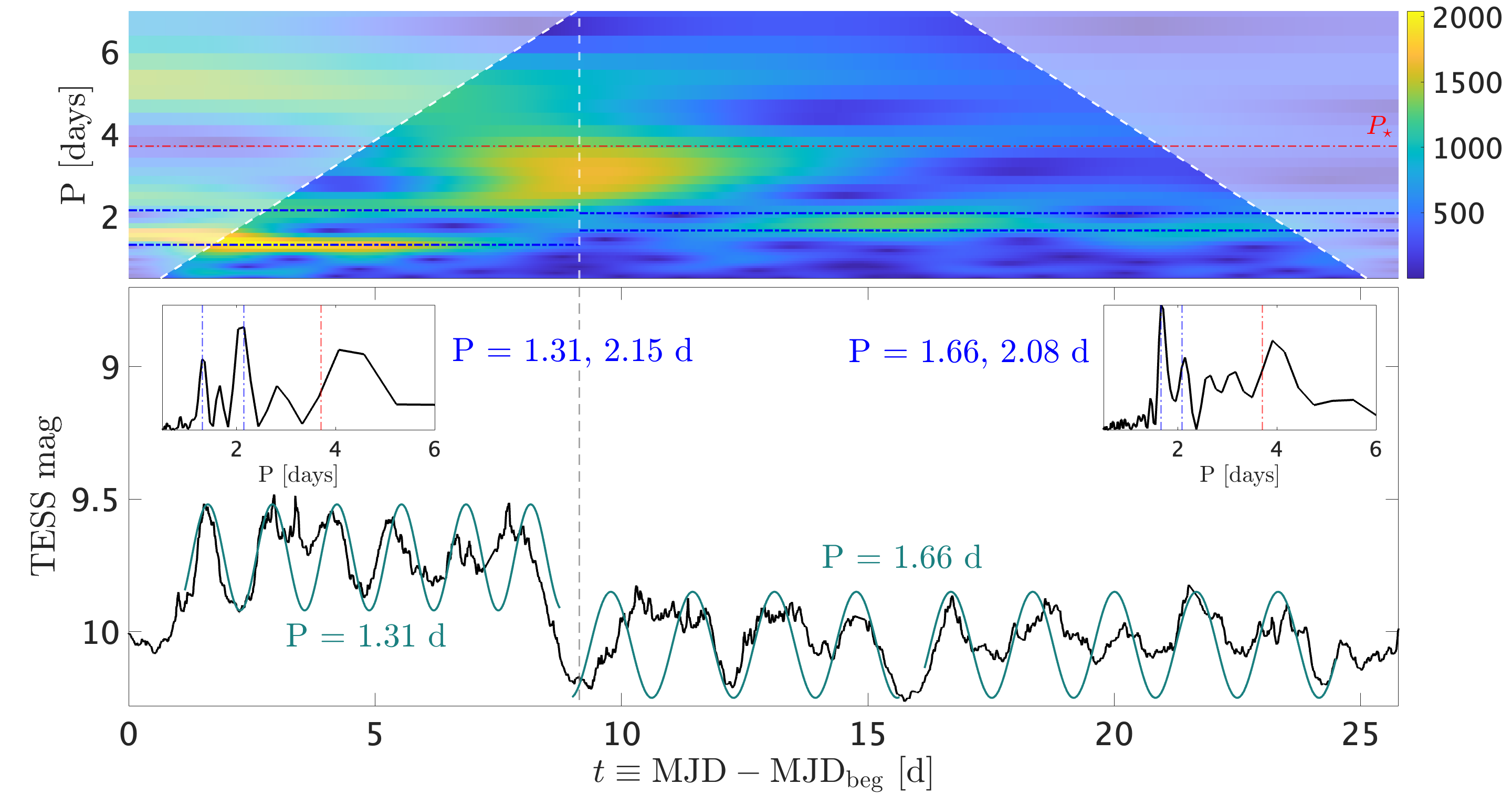}
	\caption{Continuous wavelet analysis of the TESS Sector~65 light curve. The top panel shows the CWT. The bottom panel displays the light curve. The insets show the LSP for the two portions of the light curve. The blue and red lines mark the detected period and $P_{\star}$, respectively. Sinusoidal oscillations are superposed to the TESS observation in green, with the period of the oscillations in each portion reported with the same color.}
	\label{fig:TESS_2023_CWT}
    \end{figure*}
    
    The light curve displays two epochs of quasi-periodic bursts superposed on a base level of $\sim~10.2$~mag. In the first epoch the system reaches $9.5$~mag as maximum brightness. In the second epoch this value is reduced to $9.8$~mag. The transition between these two regimes is at $t \approx 9.15$~days. A visual inspection of the light curve suggests that the typical timescale of the variability is smaller than $P_{\star}$. This can be seen, for example, between $\phi = 1.7$ and $\phi = 2.7$, where we observe three maxima instead of one. 

    We used the continuous wavelet analysis to determine the frequency content of the light curve as a function of time. % We chose as wavelet template the Morse wavelet \citep{LillyOlhede2012} with a symmetry parameter $\gamma = 3$ and a time-bandwidth product $\mathcal{P} = 45$. 
    In the analysis of the QPOs, we used the quality factor as a measure of the coherence of the oscillations. It is defined as $Q = P_0/\Delta P$, where $P_0$ is the detected period and $\Delta P$ is the FWHM of the peak.
    The continuous wavelet transform (CWT) of the light curve (Fig.~\ref{fig:TESS_2023_CWT}) reveals how these two segments have a different frequency content. To refine the estimate of the timescales, we computed a separate Lomb-Scargle Periodogram (LSP) for each segment of the light curve. For both segments, we detected a double-peaked structure in the LSP. There is power around $P_{\star}$ but the peak is broad, that is, the oscillations are not coherent.
    For $t \lesssim 9.15$~days, we obtained $P = 1.31 \pm 0.06$~days with $Q \approx 9.1$ and $P = 2.15 \pm 0.14$~days with $Q \approx 6.5$. For $t \gtrsim 9.15$~days, we found instead $P = 1.66 \pm 0.07$~days with $Q \approx 10.1$ and $P = 2.08 \pm 0.10$~days with $Q \approx 8.7$. The 1$\sigma$ uncertainties were obtained from the quality factor by converting the FWHM to the standard deviation of a Gaussian (FWHM~=~$2\sqrt{2 \ln 2} \sigma$).
    A similar multiple-peak structure of the LSP is observed in 3D MHD simulations \citep[e.g.,][]{KulkarniRomanova2009} and indicates that the oscillations are not purely sinusoidal \citep{RomanovaKulkarni2009}. % This could be due to the intrinsic variability in the shape and the brightness of the spot(s) and/or to the presence of beat frequencies \citep{RomanovaKulkarni2009}. 
    % P1 = 1.31 \pm 0.06 d and P2 = 1.66 \pm 0.07 d
    
    For both time segments, the shorter period has the higher quality factor, and we propose that this period is the actual timescale of the oscillations. We show this by superposing a sinusoidal oscillation with that period on the relative portion of the light curve. The sine function is parameterized as mag$(t) = C + A \sin [2\pi (t-t_0)/P]$. We adjusted the free parameters of the sine function (i.e., $C$, $A$, and $t_0$) in order to visually reproduce the behavior of the QPOs. To do this, we further split the second segment in two parts, with separation at $t \approx 15.75$~days. This is the time where the light curve returns to the base level, after which the oscillations seem to be delayed by $\sim 0.5$~days. Figure~\ref{fig:TESS_2023_CWT} shows the results of this procedure. Most of the maxima in the sinusoids match a peak in the TESS light curve. The cycle-to-cycle variability indicates that the environment traced by TESS changes on dynamical ($1-2$~days) timescales. We conclude that $P_1 = 1.31$~days and $P_2 = 1.66$~days are the typical timescales of the variability, while the higher period peaks in the LSPs are due to the non-sinusoidal nature of the oscillations. In the RT-unstable accretion scenario, these periods are interpreted as the Keplerian periods at the truncation radius $R_{\rm T}$.
    
    \subsection{Contemporaneous CHIRON spectroscopy}
    \label{CHIRON_TESS}
\begin{figure}
	\centering
	\includegraphics[width=\linewidth]{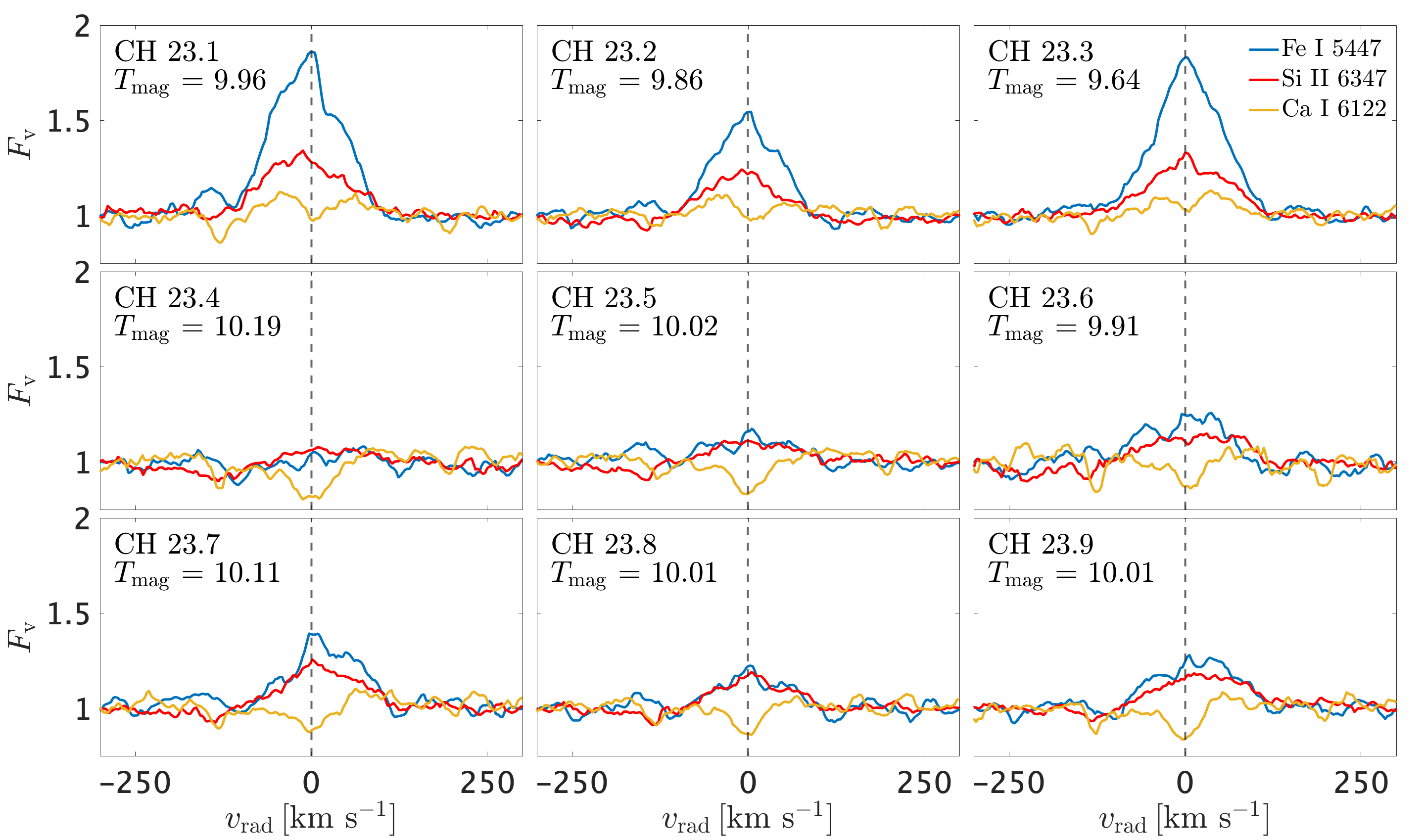}
	\caption{Normalized profiles of the \ion{Fe}{i} 5447, \ion{Si}{ii} 6347, and \ion{Ca}{i} 6122 lines in the 9 CHIRON spectra that are contemporaneous to the TESS Sector~65 light curve. $T_{\rm mag}$ is the TESS magnitude at these epochs.}
	\label{fig:BMLs_CHIRON23}
    \end{figure}

    The maximum magnitude level reached in the two segments of the TESS Sector~65 light curve differs by $\sim 0.3$~mag, that is, a factor $\sim 30\%$ in flux. The system attains the higher flux in the first segment, where the detected period is lower. 3D MHD simulations of accretion through a MBL showed how an increase in $\dot{M}_{\rm acc}$ leads to a decrease in the detected period \citep{RomanovaKulkarni2009}. If the flux increase is caused by an increase in $\dot{M}_{\rm acc}$, the change of the QPO period within the TESS Sector~65 light curve is an observational signature of $R_{\rm T}$ moving toward the star.
    The spectroscopic data from CHIRON confirm this hypothesis. 
    % The first three spectra, which are taken in the epoch of higher flux, have $\rm {VF} \gtrsim 3.85$, while for the others $\rm{VF} \lesssim 2.8$.
    Table~\ref{tab:log_specobs} shows that among the CHIRON observations from 2023 the first three spectra, which are taken in the epoch of highest flux, have the highest veiling.
    Figure~\ref{fig:BMLs_CHIRON23} shows the \ion{Fe}{i} 5447, \ion{Si}{ii} 6347, and \ion{Ca}{i} 6122 lines in the 9 CHIRON observations that are simultaneous to the TESS Sector~65 light curve. The first three spectra are the ones with the higher EW of these lines, indicating that the system is in a state of higher accretion rate in the first portion of the light curve (see the discussion in Sect.~\ref{spectroscopic_variability}).  
    % Interestingly, the EWs are not directly related to the brightness of the system in the TESS bandpass (see, e.g., CH~23.1 vs. CH~23.6).
    Assuming a constant $B_{\star}$ during the TESS observing run, the ratio $P_1/P_2$ can be converted to a ratio between accretion rates. Since $P_{\rm T} \propto R_{\rm T}^{3/2}$ and $R_{\rm T} \propto (B_{\star}^2/\dot{M}_{\rm acc})^{2/10}$ (Eq.~\ref{RT_Romanova}), we find $\dot{M}_{\rm acc, 1}/\dot{M}_{\rm acc, 2} = (P_1/P_2)^{-10/3} \approx 2.2$.
 
    \subsection{Oscillations from a nonstationary hot spot}
    \label{QPOs_origin}
    All timescales observed in the TESS light curve are shorter than the stellar rotation period. There are two regions that can produce variability at such timescales: a non-axisymmetric portion of the disk that extends inward of $R_{\rm co}$ and rotates around the star \citep{Sicilia-Aguilar+2023}, or a nonstationary hot spot on the surface of the star \citep{RomanovaKulkarni2009}. 

    A simple, planar portion of the disk which revolves around the star such as the one proposed by \citet{Sicilia-Aguilar+2023} cannot account for the quasi-periodic variability observed in RU~Lup. The reason is that the angle between the surface of the disk and the line of sight is constant, since the disk lies on a plane. 
    % Even if one takes into account the concave shape of the surface\footnote{The disk scale height $H$ is related to the radial distance $r$ as $H ~ = ~ c_{\rm s} (GM_{\star})^{-1/2} r^{3/2}$, where $c_{\rm s}$ is the sound speed.}, a sufficiently small portion of the disk can be approximated as planar. 
    % Therefore, unless the structure in the disk is eclipsed by the star, it cannot produce a rotational modulation. 
    If the structure is located in the disk, it must have a more complicated configuration, such that its projected surface changes as it revolves around the star.

\begin{figure}
        \centering
        \includegraphics[width=\linewidth]{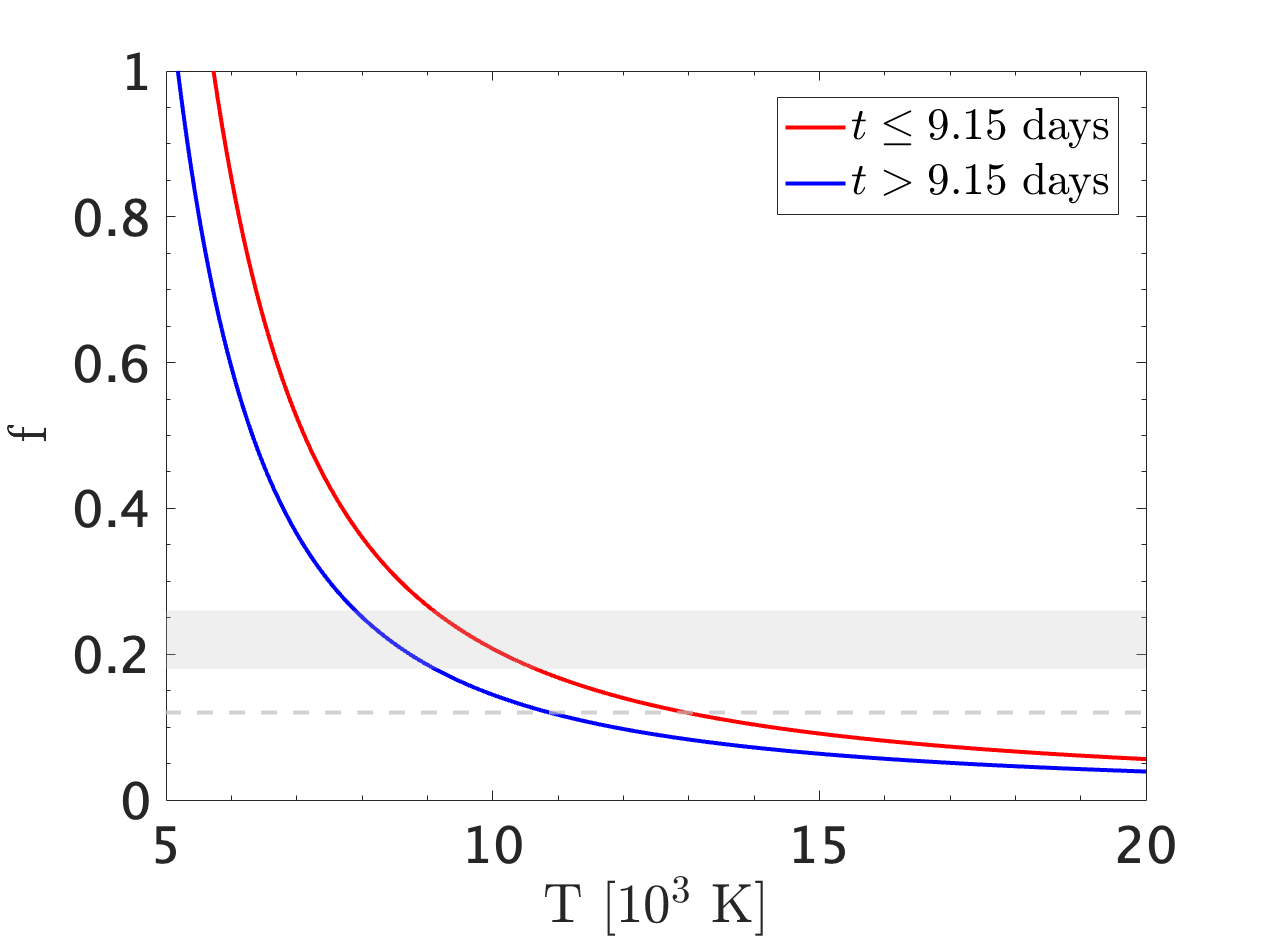}
        \caption{Relation between the filling factor ($f$) and the temperature ($T$) (Appendix~\ref{spot_model}), for a spot that reproduces the modulations observed in the TESS Sector~65 light curve. The gray area marks $0.18 \leq f \leq 0.26$ \citep{Wendeborn+2024_1}, while the dashed line marks $f = 0.12$ \citep{DodinLamzin2012}.}
        \label{fig:fvsT_spot}
    \end{figure}

    On the other hand, a moving hot spot on the surface of the star can reproduce the observed variability, since its projected area varies with the azimuthal position as the spot rotates. The amplitude and zero point of the oscillations observed with TESS can be reproduced with an analytical model, assuming a spot with temperature $T$ and filling factor $f$, located at a latitude $\theta_{\rm S}$ on the stellar surface (Appendix~\ref{spot_model}). We analyzed the two portions of the light curve with different maximum fluxes separately. We obtained $\theta_{\rm{S}1} = 61^{\rm o}$ and $\theta_{\rm{S}2} = 65^{\rm o}$ for the latitude of the spot in the two segments (Eq.~\ref{spot_latitude}). Figure~\ref{fig:fvsT_spot} shows the degenerate relation between the filling factor and the temperature of the spot that best fits the modulations in the light curve (Eq.~\ref{spot_fvsT}).
    % Although simulations of the MBL regime of accretion predict the hot spots to be close to the equator (see, e.g., Fig.~3 of \citealt{RomanovaKulkarni2009}), this seems not to be the case for RU~Lup. 
    The analytical model indicates that the spot is either more extended or hotter, or both, during the epoch of increased accretion ($t \leq 9.15$~days in Fig.~\ref{fig:fvsT_spot}).

    \section{Discussion}
    \label{discussion}
    There are two main results of this work. First, the narrow component of the \ion{He}{i}~5876 line is modulated with a period of 3.63 days, close to $P_{\star} = 3.71$ days derived by \citet{Stempels+2007} from the radial velocity changes in the absorption lines of RU~Lup. 
    % The properties of the \ion{He}{i}~5876 NC are compatible with a formation in the post-shock region, that is located close to the stellar surface.
    The fact that $v_{\rm NC}$ is in anti-phase with $v_{\rm phot}$ indicates that the photospheric lines are distorted by narrow emission components produced in a hot spot (Sect.~\ref{antiphase_RV}) and not by a cold spot as suggested by \citet{Stempels+2007}.
    
    Second, the timescales detected in the TESS light curve of RU~Lup are shorter than $P_{\star}$, and they can be interpreted as evidence of accretion from inward of $R_{\rm co}$. The photometric variability can be produced by either  a complex warped structure in the disk or spot(s) on the stellar surface. In addition, we observed a change in the period throughout the TESS light curve that appears to be related to variations in $\dot{M}_{\rm acc}$.
    In the rest of the section, we discuss scenarios that might explain the discrepancy between the timescales detected in spectroscopy and photometry. We conclude the discussion with the analysis of the accretion flow inferred from the emission lines.

    % The extensive spectroscopic and photometric database provided us with several insights into the stellar and circumstellar environment of RU~Lup.
    % We now put these results together to derive a consistent picture of the accretion structure of RU~Lup.

    \subsection{TESS light curve and regime of accretion}
    The study of the frequency spectrum of the TESS light curve allows us to obtain an indirect measure of the position of the truncation radius $R_{\rm T}$.
    Interpreting $P_1 = 1.31$~days and $P_2 = 1.66$~days (Sect.~\ref{TESS}) as the Keplerian rotation at the truncation radius $R_{\rm T}$, we convert these measures to positions of $R_{\rm T}$ using $R_{\rm T}/R_{\rm co} = (P_{\rm T}/P_{\star})^{2/3}$. We obtain $R_{\rm T} = 0.5 ~ R_{\rm co}$ for the first epoch and $R_{\rm T} = 0.59 ~ R_{\rm co}$ for the second epoch.
    The derived $R_{\rm T}/R_{\rm co}$ ratios indicate that RU Lup accretes in a RT-unstable regime \citep{Blinova+2016, Pantolmos+2020}, in agreement with the result by \citet{Stock+2022}. Together with the estimate of $R_{\rm co}$ (Table~\ref{tab:stellar_accretion_pars}), we find that $R_{\rm T}$ is approximately $2~R_{\star}$ ($\sim 1.82~R_{\star}$ and $\sim 2.15~R_{\star}$ for the first and second epoch, respectively).
    % This value is lower than the one derived by Wendeborn et al. (in prep.) for RU~Lup from a fit of the H$\alpha$, H$\beta$, and H$\gamma$ lines with an axisymmetric accretion flow model. The reason for this discrepancy might be a departure from an axisymmetric flow, in agreement with what we deduced from the analysis of the BCs (see Sect.~\ref{inner_disk_dynamics}).

    The position of the truncation radius can be estimated independently, knowing $\dot{M}_{\rm acc}$ and $B_{\star}$ (Eq.~\ref{RT_Romanova}). 
    We obtained an estimate of $\dot{M}_{\rm acc}$ for the CH~23.4 spectrum\footnote{We used this spectrum because it is the only one that is absolutely flux-calibrated among the CHIRON spectra from 2023.} using the \ion{He}{i}~5876 line as follows. We calculated the line luminosity from the integrated flux of the line using the \textit{Gaia} DR3 distance (Table~\ref{tab:stellar_accretion_pars}) and converted it to an accretion luminosity ($L_{\rm acc}$) using the relation calibrated by \citet{Alcala+2017}. Then, assuming that the free-fall starts at $R_{\rm co}$ (i.e., $R_{\rm T} = R_{\rm co}$), we derived $\dot{M}_{\rm acc}$ by inverting the relation
    \begin{equation}
        L_{\rm acc} = \frac{G M_{\star} \dot{M}_{\rm acc}}{R_{\star}} \left( 1 - \frac{R_{\star}}{R_{\rm T}} \right)
        \label{Lacc}
    \end{equation}
    \citep{Hartmann+2016}. We obtained $\dot{M}_{\rm acc} = 1.48 \times 10^{-7} ~ M_{\odot}~\rm{yr^{-1}}$. Using the upper limit of $\sim 0.5$~kG derived by \citet{JohnstonePenston1986} for the dipolar magnetic field of RU~Lup, we find an upper limit of $2.1~R_{\star}$ for $R_{\rm T}$, in good agreement with the values inferred photometrically\footnote{This derivation has the subtlety that the spectroscopic estimate of $\dot{M}_{\rm acc}$ depends in turn on $R_{\rm T}$.}.
    
    % Using $\dot{M}_{\rm acc} = 1.48 \times 10^{-7} ~ M_{\odot}~\rm{yr^{-1}}$ obtained for the ES~23.4 spectrum\footnote{We used this spectrum because it is the only one that is flux-calibrated among the CHIRON spectra from 2023.} as in Sect.~\ref{BMLs}, and the upper limit of $\sim 0.5$~kG derived by \citet{JohnstonePenston1986} for the dipolar magnetic field of RU~Lup, we find $R_{\rm T} = 2.1~R_{\star}$, in good agreement with the values inferred photometrically\footnote{This derivation has the subtlety that the spectroscopic estimate of $\dot{M}_{\rm acc}$ depends in turn on $R_{\rm T}$. In Sect.~\ref{BMLs} we assumed that $R_{\rm T} = R_{\rm co}$ (Eq.~\ref{Lacc}) to derive the accretion rate from $L_{\rm acc}$.}.
    % This leads to an underestimation of $\dot{M}_{\rm acc}$, that is, an overestimation of $R_{\rm T}$ when using Eq.~\ref{RT_Romanova}
    
    \citet{Blinova+2016} showed that the unstable regime can be divided in two sub-regimes; namely, the chaotic regime and the ordered \citep[or MBL,][]{RomanovaKulkarni2009} regime. The difference between the two regimes is in the stability of the QPOs. In the chaotic regime, short-lived hot spots, which last for only a few rotations around the star, produce short-duration QPOs \citep{KulkarniRomanova2009}. In the MBL regime, the QPOs are more stable \citep{RomanovaKulkarni2009}. The clear QPOs observed throughout the TESS Sector~65 light curve are more compatible with the latter regime of accretion. The values of $R_{\rm T}/R_{\rm co}$ that we derived above are in agreement with the results of 3D MHD simulations of accretion in the MBL regime ($R_{\rm T}/R_{\rm co} \lesssim 0.59$, \citealt{Blinova+2016}). In this scenario, accretion proceeds in two ordered tongues controlled by the RT instability, which produce hot spots on the surface of the star.

    \subsection{Properties of the hot spot inferred from spectroscopy}    
    The radial velocity amplitude of the \ion{He}{i}~5876 NC modulation is smaller than $v\sin i$, indicating that the NC-emitting region is located on the stellar surface, at high latitude. The NC emission in high energy lines (e.g., \ion{He}{i} and \ion{He}{ii}) identify this region as the post-shock region.
    
    The red asymmetry observed in the NC of the \ion{He}{i} line, with emission up to $\sim + 100 ~\rm{km~s^{-1}}$ (i.e., $\sim 3 \sigma_{\rm r}$ for \ion{He}{i}, Fig.~\ref{fig:He_metals_NC}) is compatible with formation in the post-shock region, where the gas decelerates to $1/4$ of the pre-shock velocity \citep{Hartmann+2016}. The blue wing of the line might be indicative of the broadening mechanism, with either thermal or turbulent motions that broaden the emission line. Using $\sigma = (2 k_{\rm B}T / Am_{\rm p})^{1/2}$, where $k_{\rm B}$ is the Boltzmann constant, $m_{\rm p}$ is the proton mass, and $A = 4$ is the atomic number of helium, we get an upper limit of $\sim 50000$~K for the temperature of the post-shock region from $\sigma_{\rm b} = 14 ~\rm{km~s^{-1}}$ (Fig.~\ref{fig:He_metals_NC}). The post-shock region cools in X-rays \citep{Lamzin1999, Sacco+2008} and high energy ultraviolet lines \citep{Ardila+2013}. Therefore, we do not expect it to directly contribute in the TESS bandpass. In that wavelength range, most of the contribution comes from the heated photosphere below the shock \citep{CalvetGullbring1998}. This region is heated from above by $3/4$ of the shock energy, reaching temperatures up to $\sim 8000$~K \citep{Hartmann+2016}. 
    % solved the radiative transfer equations for an atmosphere in hydrostatic balance that is heated by the external radiation of an accretion shock, and 
    
\begin{figure}
        \centering
        \includegraphics[width=\linewidth]{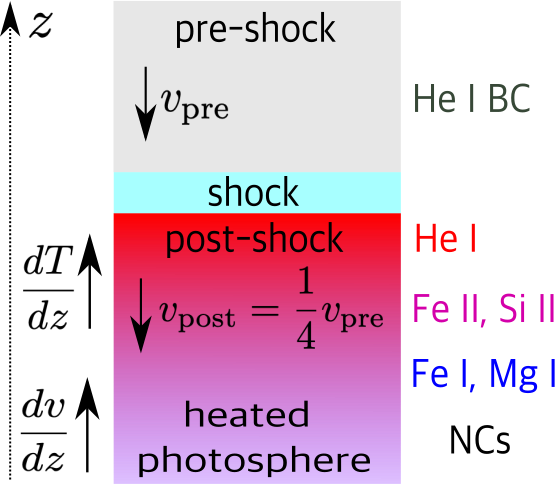}
        \caption{Sketch of the possible vertical stratification of the hot spot inferred from the analysis of the NCs. $v_{\rm pre}$ and $v_{\rm post}$ are the pre-shock and post-shock velocities.}
        \label{fig:shock_vertical_stratification}
    \end{figure}

    \citet{DodinLamzin2012} showed that the heated photosphere not only emits in the continuum, but is also responsible for the emission in the NC of the metallic species. The NCs are formed in a layer with a temperature inversion (i.e., a chromospheric-like structure) above the continuum-emitting region.
    Our analysis of the NCs possibly highlights the vertical stratification of this layer. The more energetic lines (\ion{Fe}{ii} and \ion{Si}{ii}) are indeed more asymmetric to the red than the less energetic ones (\ion{Fe}{i} and \ion{Mg}{i}), suggesting that the former could originate in the upper part of this structure (at higher T) and still have a residual infall velocity, as sketched in Fig.~\ref{fig:shock_vertical_stratification}. 

    In Sect.~\ref{veiling}, we showed how the veiling spectrum of RU~Lup at 5800~{\AA} is composed of a continuum plus line emission that fills in the photospheric absorption lines. 
    % The use of absolutely-flux calibrated spectra allowed us to directly disentangle the two components and their relative magnitude. 
    % \citet{DodinLamzin2012} claimed that these two components are formed in the heated photosphere below the shock. 
    The anticorrelation between $v_{\rm NC}$ of the \ion{He}{i}~5876 line and $v_{\rm phot}$ (Sect.~\ref{antiphase_RV}) imply that the veiling spectrum is formed in a confined region on the stellar surface, in other words, a spot. This spot, which we identify as the heated photosphere below the post-shock region, must also contribute to the emission in the TESS bandpass, which extends from 5800 to 11200~{\AA}.

    \subsection{Properties of the hot spot inferred from photometry}
    \label{spot_props}
    By using the analytical model of Appendix~\ref{spot_model}, we were able to infer the properties of the photometric spot, assuming that it is nonstationary on the surface of the star.
    The degeneracy between the filling factor and the temperature of the spot can be removed by considering values of $f$ from the literature. Using a typical value of $f \approx 0.05$ for CTTSs \citep{CalvetGullbring1998, Calvet+2004}, we get a temperature of $\sim 20000$~K for the spot. However, specific works on RU~Lup showed how the filling factor of the spot is higher than the usual values for CTTSs. In Fig.~\ref{fig:fvsT_spot} we report the values obtained by \citet{Wendeborn+2024_1} from a best fit of the 2021 ultraviolet spectra of RU~Lup with a shock model ($0.18 \leq f \leq 0.26$) and the value derived by \citet{DodinLamzin2012} by fitting the veiling spectrum of RU Lup ($f = 0.12$). In these cases, we obtain a colder spot with $T \sim 8000-12000$~K, more compatible with the typical values for CTTSs \citep{Hartmann+2016}.

    Assuming a spherical cap geometry for the spot (Appendix~\ref{f_cap}), and a filling factor between 0.1 and 0.2 as mentioned in the previous literature of RU~Lup, the half-opening angle of the spot is between $37^{\rm o}$ and $57^{\rm o}$. If the hot spot is as extended as that, then the spot model from Appendix~\ref{spot_model} is too simplistic and the derived $\theta_{\rm S}$ must be regarded as an average latitude of the continuum-emitting region. 
    % This is a limitation of the model, which assumes a hot spot that is located on a point on the stellar surface and emits a flux equal to $4\pi R_{\star}^2 f B_{\lambda}(T)$, where $B_{\lambda}(T)$ is the Plank function. 

    \subsection{Comparison between spectroscopy, photometry, and magnetohydrodynamic simulations of the spot}
    \label{spec_vs_sim_vs_phot}
    The spectroscopic results on the hot spot seem to disagree with what is derived from the analysis of the TESS light curve and with the results of 3D MHD simulations of a system accreting in the MBL regime, due to the latitude/extension of the spots and the detected timescales. 
    We propose that the spectroscopic and photometric spot are part of the same structure, which we identify as the region in which the accreting gas impacts the stellar surface. The \ion{He}{i}~NC — that is, the tracer of the spectroscopic spot — originates in the post-shock region. The continuum emission in the TESS bandpass, associated with the photometric spot, is formed instead in the heated photosphere below the shock. This region emits also in the NC of the metallic species and it is responsible for the line-filling component of the veiling \citep{DodinLamzin2012}.
    
    Considering the typical extension of the spot in RU~Lup ($f \approx 0.1-0.2$), we might expect a discrepancy between the latitude derived from spectroscopy and photometry. The region traced by the \ion{He}{i} NC (i.e., the post-shock region) must be somewhat less extended than the heated photosphere traced by TESS. Otherwise, considering the low inclination of the system and the intrinsic variability of this region as a consequence of the accretion process, it would be impossible to detect a rotational modulation \citep{Sicilia-Aguilar+2023}. 

    A possible explanation of the inconsistency in the detected periods lies in the variability of the mass accretion rate.
    The $3.63$~day period was derived in Sect.~\ref{HeI5876_NC} using data from epochs that have different $\dot{M}_{\rm acc}$. In Sect.~\ref{TESS} we discussed how variations in the accretion rate lead to variations in the detected period. 
    \citet{Stock+2022} showed that the accretion rate of RU~Lup varies by a factor of $\sim 2$ on a timescale of weeks. Since $P_{\rm T} \propto \dot{M}_{\rm acc}^{-3/10}$ (Eq.~\ref{RT_Romanova}), this means that the period of a nonstationary hot spot would vary by a factor of $2^{3/10} \approx 1.23$ on such timescales.
    If the \ion{He}{i}~5876 NC traces a nonstationary hot spot on the stellar surface produced by tongues originating at $R_{\rm T}$, it is plausible that its radial velocity is modulated with both $P_{\star}$ and the inner disk period ($P_{\rm T}$). In this situation, if one were to look for periodicity combining observations that represent different accretion states, the power at $P_{\star}$ would be enhanced in the periodogram because of the stability of this period among different epochs.

\begin{figure}
        \centering
        \includegraphics[width=\linewidth]{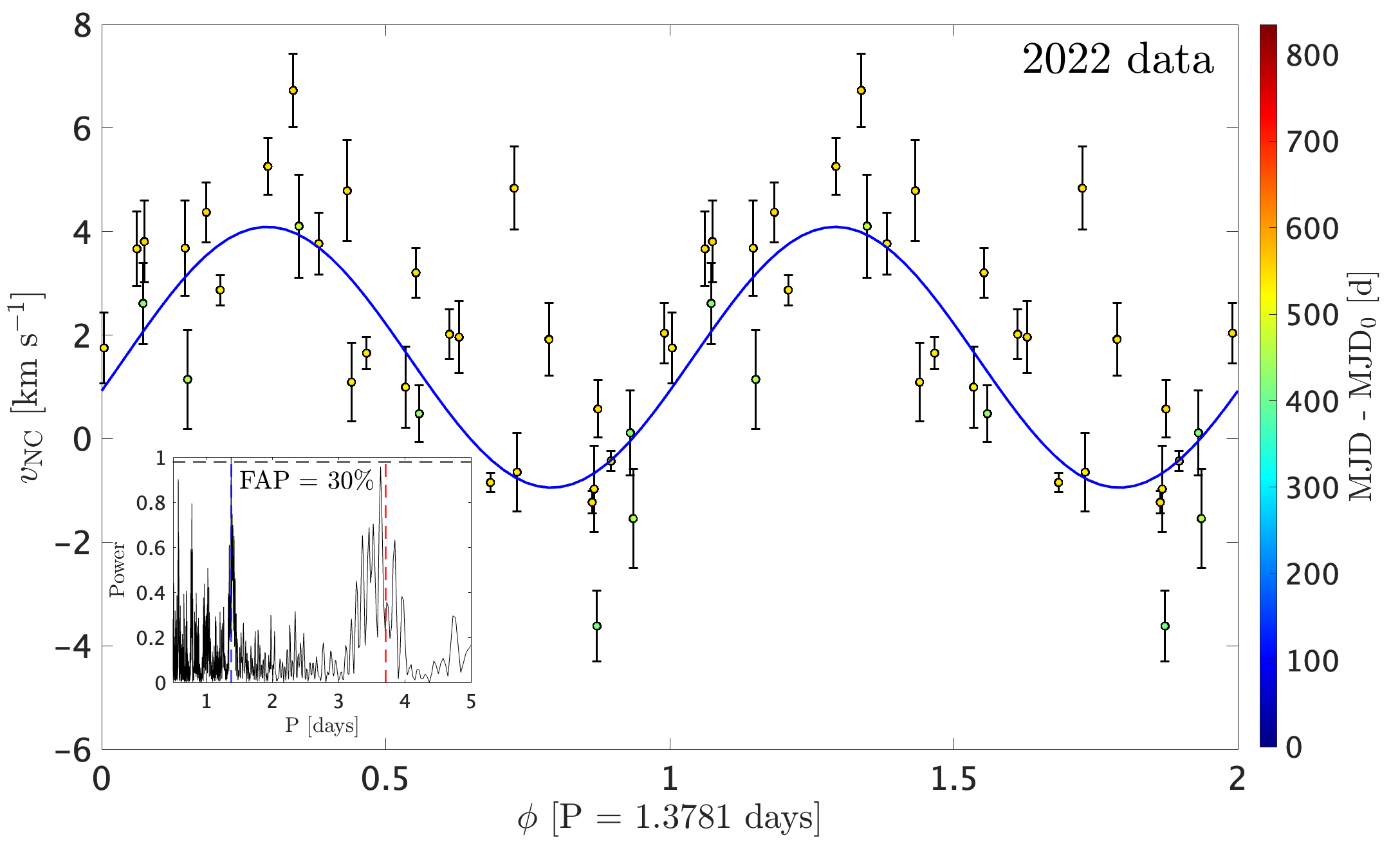}
        \caption{Radial velocity curve of the \ion{He}{i}~5876 NC for the 2022 observations only. The inset shows the LSP, with the detected $P$ and $P_{\star}$ marked with dashed blue and red lines, respectively.}
        \label{fig:HeI_5876_2022}
    \end{figure}

    These complications are illustrated in Fig.~\ref{fig:HeI_5876_2022}, where we recomputed the LSP of the \ion{He}{i}~5876 NC using only the data from 2022. The periodogram has maximum power at $P = 1.38$~days, but the FAP of this signal is $> 30\%$; that is, it is not statistically significant. The second highest peak is close to $P_{\star}$. 
    The period with maximum power is similar to the periods derived from TESS. This suggests that the apparent lack of continuum emission from the spot (Sect.~\ref{TESS}) could be actually caused by an incorrect phase-folding of the radial velocity of the \ion{He}{I}~5876 NC. If the period of the hot spot is not correct, then the phase $\phi_S$ where we expect its maximum contribution is different from what we derived, and might be in agreement with maxima in the TESS light curve.

    Therefore, we conclude that in the MBL picture the disagreement between the period detected from spectroscopy ($P_{\star}$) and the period inferred from photometry ($< P_{\star}$) might be attributed to an observational problem, namely, the difficulty of tracing a hot spot that rotates with the inner disk period, with the latter being sensitive to variations in $\dot{M}_{\rm acc}$.  

    \subsection{Alternative explanations: sub-structure of the spot and inner disk warp}
    Although accretion in the MBL regime can explain the photometric behavior of RU~Lup, the detected periods are close to $P_{\star}/2$. 
    Hence, the photometric behavior could also be explained with a modulation at a period close to $P_{\star}$ but with a brightness distribution on the visible hemisphere more closely resembling two hot spots; that is, a single, extended hot spot that has a non-homogeneous surface brightness, such that we see two brighter features. The brightness distribution of this region could vary on dynamical timescales and the detected periods could be aliases produced by the complex structure of this region. Similarly, the change in the quasi-period between the two epochs of the TESS light curve could be due to an intrinsic variability in the shape and brightness of this extended spot. Since the velocity modulation of emission lines mainly traces the average brightness distribution, a large non-homogeneous spot would lead to a modulation at $\sim P_{\star}$ in the radial velocity of the NCs, as we observe for the \ion{He}{i} 5876 line.

    % Indeed, assuming that the period detected in the first portion of the TESS light curve, $P_1 = 1.31$~days, is half of the period at $R_{\rm T}$, we would obtain $R_{\rm T}/R_{\rm co} = (2P_1/P_{\star})^{2/3} = 0.79$. This value is higher than 0.71, that is, the limit for the onset of unstable accretion according to \citet{Blinova+2016}.
    
    Periods close to $P_{\star}$ would be compatible with a scenario in which RU Lup is in a stable accretion regime. However, 3D MHD simulations of accretion in the stable regime showed how the spots are formed close to the magnetic pole and are antipodal \citep[e.g.,][]{Romanova+2003, Romanova+2004} so that it is unlikely to see two distinct, antipodal spots given the low inclination ($i_{\star} = 16 \pm 6~{}^{\rm o}$) of the stellar rotation axis in RU Lup. Rather, the brightness modulation would require sub-structure of a single spot.
    MHD simulations of accretion in the stable regime predict the hot spots to be bow-shaped around the magnetic axis \citep{Romanova+2004} with a temperature gradient from the center to the edges \citep{KulkarniRomanova2013}; that is, a structure that is different from the requirements of period aliasing. Hence, the structure of the hot spot must be somewhat different from what is predicted by 3D MHD simulations if accretion in RU Lup proceeds in the stable regime.
    Moreover, independent of the true variability, the TESS light curve shows that the dominant periodicity changes, with the shorter period pertaining to epochs with higher (optical) luminosity, in agreement with MHD models of accretion in the unstable regime \citep{KulkarniRomanova2009}.
    Therefore, we think that the MBL scenario fits the observations better than the hypothesis of accretion in a stable regime. 

    A third explanation is motivated by the high values of the filling factor that we derived for RU~Lup. Such a high (0.1-0.2) filling factor could be achieved if the part of the continuum-emitting region lies in the disk. In Sect.~\ref{QPOs_origin} we discussed how a structure located in the disk must have a complex geometry in order to produce a quasi-periodic modulation. A warp in the inner disk inward of $R_{\rm co}$ might satisfy this condition, and would naturally explain the observed timescales. 
    % Assuming that the warp extends from $R_{\rm T} \approx 2 ~ R_{\star}$ to the stellar surface, the filling factor would be
    %\begin{equation}
    %    f = \frac{1}{4 \pi R_{\star}^2} \int_{R_{\star}}^{R_{\rm T}} \int_{0}^{\Delta \phi} r dr d\phi = \frac{1}{8 \pi R_{\star}^2} (R_{\rm T}^2 - R_{\star}^2) \Delta \phi \approx \frac{3 \Delta \phi}{8\pi}
    % \end{equation}
    % A filling factor of $0.2$ could be achieved if the warp has an azimuthal extension of $\sim 96^{\rm o}$.
    In this scenario, the continuum emitting region would be spatially separated from the region emitting in the NCs, because the latter is unequivocally associated with shocked gas on the stellar surface. Such a structure could be characteristic of systems accreting through a compact magnetosphere, and it might be a non-axisymmetric version of the classical boundary layer. To our knowledge, no models of quasi periodic oscillations from a warped structure in the disk have been studied.

    \subsection{Inner disk dynamics}
    \label{inner_disk_dynamics}
    
    \subsubsection{Structure of the flow}
    The width of the metallic lines points toward an origin in the circumstellar matter.
    Their velocity centroids ($v_{\rm BC}$) are between $-30$ and $+60 ~ \rm{km~s^{-1}}$ (Fig.~\ref{fig:BMLs_vBC_vs_LA}),  compatible with the projected velocities of flows within the inner disk. For a Keplerian disk seen at an inclination $i$, the radial velocity is
    \begin{equation}
        v_{\rm rad}(r, \phi_{\rm D}) = \sqrt{\frac{GM_{\star}}{r}} \sin i \sin \phi_{\rm D}
    \end{equation}
    where $r$ is the distance from the star and $\phi_{\rm D}$ is the azimuth relative to the observer \citep[e.g.,][]{HorneMarsh1986}. For $i = 16~{}^{\rm o}$ (Table~\ref{tab:stellar_accretion_pars}), we get a maximum $v_{\rm rad}$ of $\sim 50, 45, 40$, and $30 ~ \rm{km~s^{-1}}$ for $r = 0.4, 0.5, 0.6$, and $1 ~ R_{\rm{co}}$. Hence, the observed $v_{\rm BC}$ values are compatible with the position of the truncation radius derived from TESS data, suggesting that the lines are formed in a disk-like structure that revolves around the star. 
    
    % However, the line-emitting region must be more complicated than a Keplerian disk.
    To reproduce the variability of the line asymmetry, this structure must be non-axisymmetric, and it might look like a portion of a Keplerian disk, as proposed by \citet{Sicilia-Aguilar+2023} to model the BC variability of the \ion{Ca}{ii} lines in EX~Lup.
    % This is in agreement with the simulations of accretion in the unstable regime \citep[e.g.,][]{KulkarniRomanova2008} and with what is observed in other strong accretors, such as RW~Aur \citep{Petrov+2001} and EX~Lup \citep{Sicilia-Aguilar+2023}. 
    However, the lines are much broader than the maximum $v_{\rm rad}$ that can be observed from a Keplerian disk, that is, $(GM_{\star}/R_{\star})^{1/2} \sin i \approx 60 ~ \rm{km~s^{-1}}$ for RU~Lup. Therefore, they are either broadened by turbulent motions in the Keplerian portion of the disk, as proposed by \citet{Sicilia-Aguilar+2023}, or formed in a flow with higher velocities.
    % with emission up to $\sim 150 ~ \rm{km~s^{-1}}$ for the \ion{Fe}{i}~5447 line,
    % The emission lines from an axisymmetric Keplerian disk are double-peaked, with peaks at the Keplerian velocity at the outer boundary of the region \citep{HorneMarsh1986}. The double-peaked structure is not observed in any of the BMLs. Since it is unlikely that the emitting region extends very far from the star, the flow is likely not confined to a disk structure.
    
    % \citet{Sicilia-Aguilar+2023} modelled the variability of the \ion{Ca}{ii} infrared lines in EX~Lup with an irradiated Keplerian disk that has a limited azimuthal extension. The model is able to reproduce the variability of the line asymmetry, although the authors had to include a turbulent broadening of 5 times the thermal velocity to fit the line width.
    In the case of RU~Lup, the analysis of the \ion{Fe}{ii}~5317 line (Fig.~\ref{fig:FeII_ES22_subtraction}) reveals how a turbulent portion of a Keplerian disk cannot completely account for the line velocities. The discrete emission components detected with the subtraction of consecutive spectra have centroids — bulk velocities $> (GM_{\star}/R_{\star})^{1/2} \sin i$ — indicating that part of the line emission is produced by macroscopic flows in non-Keplerian motion.
    
    Radial velocities of the order of $100 - 150 ~ \rm{km~s^{-1}}$ are compatible with free fall along dipolar magnetic field lines. The free fall velocity starting from rest at $R_{\rm T}$ is
    \begin{equation}
        v_{\rm ff}(r) = \sqrt{GM_{\star} ~ \left( \frac{1}{r} - \frac{1}{R_{\rm T}} \right)}.
    \end{equation}
    For $R_{\rm T} = 2 ~ R_{\star}$ (Sect.~\ref{TESS}) we get $v_{\rm ff}(R_{\star}) \approx 205 ~ \rm{km~s^{-1}}$. The observed velocity is significantly lower for a pole-on system, because the velocity vector of the gas moving along the magnetic field lines is almost transverse to the line of sight. The observed radial velocity is reduced by a factor 
    \begin{equation}
        \frac{(3/2) \sin 2 \theta  \cos \phi \sin i + (2 - 3\cos^2 \theta) \cos i}{\sqrt{4 - 3\cos^2 \theta}}
    \end{equation}
    \citep{CalvetHartmann1992, Wilson+2022}. Here, $\theta$ is the latitude, $\phi$ is the azimuth relative to the observer, and the gas is assumed to be moving along a dipolar streamline with equation $r~=~R_{\rm T} \cos^2 \theta$. The maximum redshifted and blueshifted radial velocities that can be observed are obtained for $\phi = 0$ and $\pi$ and are $+110$ and $-60 ~ \rm{km~s^{-1}}$, respectively, assuming $R_{\rm T} = 2 ~ R_{\star}$. The fact that the maximum negative $v_{\rm rad}$ is lower than maximum positive $v_{\rm rad}$ is a projection effect. This red-blue asymmetry is similar to the variation in $v_{\rm BC}$ for RU~Lup throughout our observations. This argues in favor of line formation in a magnetospheric accretion column and against formation in a structure that is confined to a plane, such as a Keplerian disk or equatorial accretion tongues. In that case, the redshifted and blueshifted line shifts would be the same.

    \subsubsection{Temperature stratification}
    The different line widths, which correlate with the excitation energy of the analyzed emission lines (Sect.~\ref{spectroscopic_variability}), indicate the existence of a stratification in the accretion flow of RU~Lup. The \ion{He}{i} lines must be formed close to the shock in high temperature conditions, and their higher FWHM relative to the \ion{Fe}{i} and \ion{Si}{ii} lines indicate a formation in a flow with higher velocities. Conversely, neutral species like \ion{Fe}{i} and \ion{Ca}{i} are formed closer to the disk, where the material starts to accrete onto the star in accretion tongues. This region must be non-axisymmetric in order to explain the line variability, and it might be similar to the azimuthally-limited Keplerian disk proposed by \citet{Sicilia-Aguilar+2023}. 
    % Assuming for simplicity a flat, irradiated disk, the temperature $T_{\rm D}$ at a distance $r$ from the star is
    % \begin{equation}
    %    T_{\rm D}(r) = \left(\frac{2}{3\pi}\right)^{1/4} \left(\frac{R_{\star}}{r}\right)^{3/4} T_{\rm eff}
    % \end{equation}
    % \citep{ChiangGoldreich1997}. For $r = 2 ~ R_{\star}$, we get $T_{\rm D} ~ \approx ~ 1700$~K. % This temperature is compatible with the energies required to excite the lines of neutral species.
    The singly ionized species such as \ion{Fe}{ii} and \ion{Si}{ii} have velocities that are not consistent with Keplerian flows, and require higher temperature conditions than what can be achieved in an irradiated disk. The extreme line asymmetry to the red that is sometimes observed in the \ion{Si}{ii}~6347 line (see Fig.~\ref{fig:BMLs_examples}) suggests that these species are partly formed in the flow along the magnetic field lines.
    The different line asymmetry observed for emission lines from the same observation support the stratification hypothesis, since lines tracing different regions of the accretion flow might be asymmetric at different times.
    
    % According to our picture, if a parcel of gas starts accreting from the disk to the star, we should first see the \ion{Fe}{i} line becoming asymmetric, followed by the \ion{Si}{ii} and the \ion{He}{i} lines. For a velocity of $100 ~ \rm{km~s^{-1}} \approx 12$~$R_{\sun}~\rm{d^{-1}}$, the gas takes $\sim 5$~hours to travel a distance of $1 ~ R_{\star}$. Therefore, lines tracing different regions of the accretion flow might be asymmetric at different times.

    In conclusion, the variability of the \ion{He}{i} and metallic lines can be explained with the lines being formed in a temperature stratified structure which is a combination of a non-axisymmetric Keplerian disk and an inflow along magnetic field lines. This structure is sensitive to variations in the accretion rate. 
    % The velocity structure of the accretion flow must be more complicated. This is expected if the system accretes through a MBL. 

    \subsection{Formation of the metallic lines}
    \begin{figure}
        \centering
        \includegraphics[width=\linewidth]{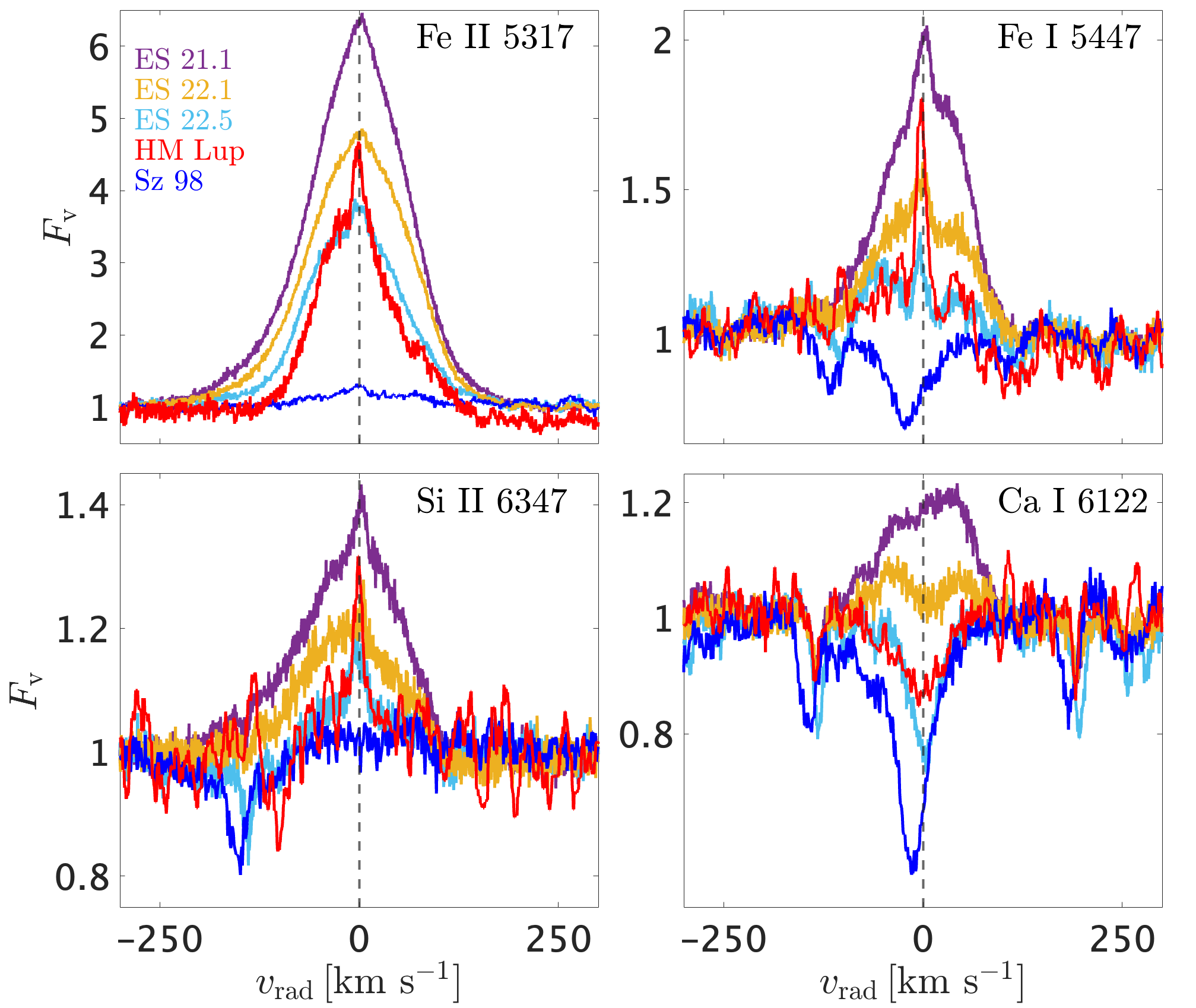}
        \caption{A selection of metallic lines in the spectra of RU~Lup, HM~Lup, and Sz 98.}
        \label{fig:BMLs_RU_HM_Lup_Sz98}
    \end{figure}
    Analogous to the case of the \ion{H}{i}, \ion{He}{i}, and \ion{Ca}{ii} lines \citep{Alcala+2017}, one would expect the strength of the metallic lines to be dependent on $\dot{M}_{\rm acc}$.
    % Since $R_{\rm T}\propto \dot{M}_{\rm acc}^{-2/10}$, this would suggest that the closer $R_{\rm T}$ is to the star, the stronger these lines are.
    Figure~\ref{fig:BMLs_RU_HM_Lup_Sz98} compares RU~Lup with two other CTTSs; namely HM~Lup, which shows prominent metallic emission, and Sz~98, which has the same stellar parameters of RU~Lup \citep{Manara+2023}. We show the second ESPRESSO observation analyzed by \citet{Armeni+2023} for HM~Lup, and an Ultraviolet and Visual Echelle Spectrograph \citep[UVES,][]{Dekker+2000} spectrum taken on 10/05/2022 as part of the PENELLOPE survey for Sz~98.
    The accretion rates for HM~Lup and Sz~98, taken from the literature, are $0.95 \times 10^{-8} ~ M_{\odot} ~ \rm{yr^{-1}}$ \citep{Armeni+2023} and $3.6 \times 10^{-8} ~ M_{\odot} ~ \rm{yr^{-1}}$ \citep{Manara+2023}, respectively. 

    The comparison between the spectrum of HM~Lup and the ES~22.5 spectrum of RU~Lup shows that despite $\dot{M}_{\rm acc}$ being one order of magnitude lower for HM~Lup, the emission line profiles are very similar. The only strong emission line is \ion{Fe}{ii}~5317. The \ion{Fe}{i}~5447 and \ion{Si}{ii}~6347 lines are weak, while the \ion{Ca}{i}~6122 is in absorption in both spectra. 
    Although Sz~98 has $\dot{M}_{\rm acc}$ between HM~Lup and RU~Lup, Fig.~\ref{fig:BMLs_RU_HM_Lup_Sz98} shows how the metallic lines are very weak in its spectrum. The only detectable emission is in \ion{Fe}{ii}~5317, with the line being slightly above the continuum.
    This suggests that the strength of these transitions relative to the continuum is not directly related to $\dot{M}_{\rm acc}$. We propose that the actual parameter that controls the strength of the metallic lines is the ratio $R_{\rm T}/R_{\rm co}$. The smaller this ratio is, the stronger the emission lines are. Different species appear at different $R_{\rm T}/R_{\rm co}$ ratios, with the \ion{Fe}{ii} lines being the first ones that show up, followed by the \ion{Fe}{i} and \ion{Si}{ii} lines, and, finally, from the \ion{Ca}{i} $\lambda\lambda~6103, 6122$ doublet. This transition is visually illustrated by the ES~22.5, ES~22.2 and ES~21.1 spectra in Fig~\ref{fig:BMLs_RU_HM_Lup_Sz98}.

    % We can compare these spectra to, e.g., the CH~23.4 observation, for which we estimated $R_{\rm T} = 0.59 ~ R_{\rm co}$ from the TESS Sector~65 light curve. In that spectrum, emission in the \ion{Fe}{i}~5447 and \ion{Si}{ii}~6347 lines is absent, and the \ion{Ca}{i} line is still in absorption. This suggests that the disk is much closer to the star during the ESPRESSO observations from 2021. 
    % as a reference for the ratio $R_{\rm T}/R_{\rm co}$. Then, from the scaling $R_{\rm T} \propto M_{\rm acc}^{-2/10}$, we estimate 

    If related to $R_{\rm T}/R_{\rm co}$, emission in metallic species is a direct indication of accretion in the RT-unstable regime. This can be explained by a simple energetic argument. When $R_{\rm T} < R_{\rm co}$, the material at $R_{\rm T}$ rotates faster than the star. Since the magnetic field lines co-rotate with the star, the gas must dissipate energy and angular momentum in order to accrete. This dissipation could give rise to the local heating required to collisionally excite the metallic lines \citep{Beristain+1998}.
  
    \section{Conclusions}
    \label{conclusions}
\begin{figure}
        \centering
        \includegraphics[width=\linewidth]{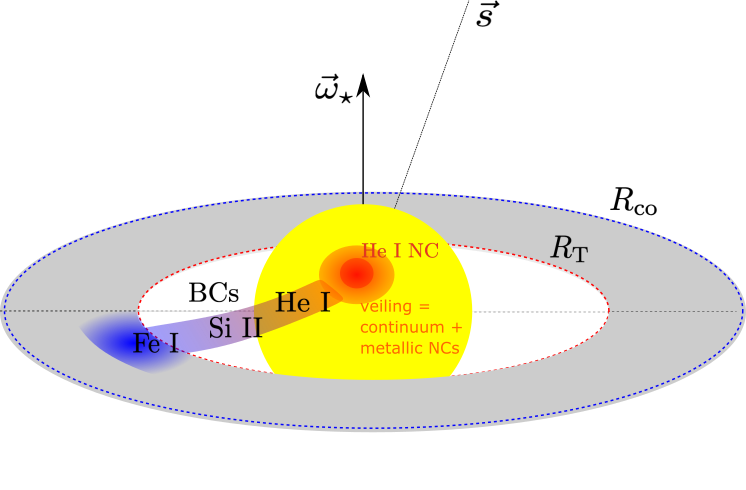}
        \caption{Schematic picture of the circumstellar environment of RU~Lup. $\vec{\omega}_{\star}$ and $\vec{\hat{s}}$ are the stellar rotation axis and the line of sight to the observer, respectively.}
        \label{fig:RULup_schematic}
    \end{figure}

    We have presented a spectrophotometric study of the CTTS RU~Lup. Figure~\ref{fig:RULup_schematic} shows a schematic picture of the circumstellar environment of RU~Lup that we inferred from our observations. The main results are the following.
    \begin{itemize}
        \item We have improved the measurement of the stellar parameters, summarized in Table~\ref{tab:stellar_accretion_pars}.
        \item We detected a modulation at a period of $3.63$~days, close to $P_{\star} = 3.71$~days \citep{Stempels+2007}, in the NC of the \ion{He}{i}~5876 line, indicating the presence of a compact region on the stellar surface that we identify as the post-shock region that originates at the footprint of the magnetic field.
        \item The heated photosphere below the accretion shock (i.e., a hot spot) is also responsible for the veiling of the photospheric lines in the spectrum.
        \item This veiling spectrum is composed of a continuum component plus line emission that fills in the stellar absorption lines. The use of flux-calibrated, high-resolution spectra allowed us to disentangle these two contributions.
        \item We detected timescales shorter than $P_{\star}$ in the TESS light curve of RU~Lup. These timescales are related to the Keplerian period at the truncation radius, $R_{\rm T}$.
        \item From these timescales, we inferred the size of the magnetosphere of RU~Lup to be $\sim 2 ~ R_{\star}$. 
        \item The behavior of the TESS Sector~65 light curve of RU~Lup is consistent with simulations of accretion through a MBL \citep{RomanovaKulkarni2009}, in which the QPOs are produced by a non-stationary hot spot on the stellar surface. 
        \item The hot spot is not as equatorial as predicted by the MHD simulations of accretion in the MBL regime, and it appears to be more extended than in typical CTTSs, with a filling factor of $\sim 0.1-0.2$.
        \item Alternatively, more complex explanations would require either a spot with a complex shape, perhaps made of two brighter knots which vary on dynamical timescales, or a warped structure in the inner disk.
        \item The BCs are formed in a non-axisymmetric, temperature stratified flow around the star in which the gas leaves the accretion disk at $R_{\rm T}$ and accretes onto the star in tongues of matter channeled by the magnetic field.
        \item The strength of the metallic emission lines might be an indicator of accretion in the RT-unstable regime, being related to the ratio between $R_{\rm T}$ and $R_{\rm co}$.   
    \end{itemize}
    
    The spectrophotometric analysis presented in this work revealed the complexity of the accretion process in RU~Lup, and showed how the physics regulating the accretion flow in this system might be somewhat different from the current theoretical paradigms of accreting CTTSs. Future studies should aim at covering high-cadence photometric data (TESS) with high-resolution spectroscopic observations, in order to try to infer the structure of the spot. An update of the measurement of the stellar magnetic field through spectro-polarimetry would be needed to more precisely derive the position of the truncation radius and confirm or reject the hypothesis of accretion through a compact magnetosphere.
   
    \bigskip

    \begin{acknowledgements} 
        The authors thank the anonymous referee for their review of this manuscript.
        This work has been supported by Deutsche Forschungsgemeinschaft (DFG) in the framework of the YTTHACA Project (469334657) under the project codes STE 1068/9-1 and MA 8447/1-1.
        PCS acknowledges support from DLR 50 OR 2205.
        AF, EF, and JMA acknowledge financial support from the project PRIN-INAF 2019 "Spectroscopically Tracing the Disk Dispersal Evolution" (STRADE) and the Large Grant INAF 2022 "YSOs Outflows, Disks and Accretion: towards a global framework for the evolution of planet forming systems" (YODA).
        CFM and JCW are funded by the European Union (ERC, WANDA, 101039452). Views and opinions expressed are however those of the author(s) only and do not necessarily reflect those of the European Union or the European Research Council Executive Agency. Neither the European Union nor the granting authority can be held responsible for them. 
        JFG was supported by Fundação para a Ciência e Tecnologia (FCT) through the research grants UIDB/04434/2020 and UIDP/04434/2020.
        This work benefited from discussions with the ODYSSEUS team\footnote{\url{https://sites.bu.edu/odysseus/}} (HST AR-16129).
        Observing time with Chiron was awarded through NOIRLab programs 2022a-492217, 2022B-994458, and 2023A-579477 (PI FMW). Chiron is operated by the SMARTS Consortium.
        Funding for the TESS mission is provided by NASA’s Science Mission directorate.
        The authors acknowledge Thomas Sperling, Michael Siwak, Ignacio Mendigutía, Konstantin Grankin, Rebeca Garcia López, and Jerome Bouvier for comments and suggestions on this work.
        The authors acknowledge with thanks the variable star observations from the {\it AAVSO International Database} contributed by observers worldwide and used in this research, and Elizabeth Waagen for coordinating the AAVSO Alerts.
        The authors acknowledge the use of the electronic bibliography maintained by the NASA/ADS\footnote{\url{https://ui.adsabs.harvard.edu}} system.
    \end{acknowledgements}

    \bigskip

    \bibliographystyle{aa} % style aa.bst 
    \bibliography{RU_LUP.bib} % your references Yourfile.bib
	
    \begin{appendix}

        \section{Log of spectroscopic observations}
        Table~\ref{tab:log_specobs} reports the log of the spectroscopic observations.
        
        \begin{table*}
            \centering
        	\caption{Log of the spectroscopic observations.} 
        	\begin{tabular}{ccccc|ccccc}
                \hline
                ID & MJD & S/N & VF & V$_{\rm mag}$ & ID & MJD & S/N & VF & V$_{\rm mag}$ \\
                 & (days) & & & (mag) & & (days) & & & (mag) \\
                \hline
                CH 21.1 & 0.04 & 65  & $5.4 \pm 0.9^{(\ddagger)}$ & - & CH 22.12 & 537.64 & 50 & $2.9 \pm 0.4$ & 10.79 \\
                CH 21.2 & 53.92 & 66 & $12 \pm 2$ & - & ES 22.2 & 537.73 & 44 & $3.2 \pm 0.4^{(\dagger)}$ & 11.14 \\
                CH 21.3 & 130.72 & 58 & - & 10.78 & CH 22.13 & 538.63 & 44 & $3.5 \pm 0.5$ & 11.22 \\
                CH 21.4 & 170.66 & 29 & - & 10.73 & CH 22.14 & 539.66 & 45 & $2.8 \pm 0.6^{(\ddagger)}$ & 11.29 \\
                CH 21.5 & 171.64 & 51 & $4.1 \pm 0.5$ & 10.89 & ES 22.3 & 539.76 & 42 & $2.9 \pm 0.4$ & 11.29 \\
                CH 21.6 & 172.66 & 70 & $9 \pm 1^{(\dagger)}$ & - & CH 22.15 & 540.72 & 65 & $4.0 \pm 0.5$ & 11.07 \\
                CH 21.7 & 173.64 & 63 & $11 \pm 2^{(\dagger)}$ & - & CH 22.16 & 541.66 & 44 & $2.4 \pm 0.4$ & 11.39 \\
                CH 21.8 & 174.68 & 65 & $6.5 \pm 0.7$ & 10.39 & CH 22.17 & 542.66 & 48 & $1.4 \pm 0.3$ & 11.74 \\
                CH 21.9 & 175.65 & 52 & $5.3 \pm 0.6$ & 10.78 & ES 22.4 & 542.76 & 24 & $1.8 \pm 0.2$ & 11.74 \\
                CH 21.10 & 177.70 & 44 & - & 10.68 & CH 22.18 & 543.71 & 42 & $1.8 \pm 0.4$ & 11.81 \\
                CH 21.11 & 178.68 & 52 & $8 \pm 1^{(\dagger)}$ & 10.63 & CH 22.19 & 544.73 & 50 & $1.9 \pm 0.4$ & 11.59 \\
                CH 21.12 & 178.80 & 62 & - & 10.63 & CH 22.20 & 545.70 & 40 & $1.9 \pm 0.4$ & 11.63 \\
                CH 21.13 & 179.68 & 57 & $10 \pm 1$ & 10.56 & CH 22.21 & 546.71 & 50 & $3.9 \pm 0.5$ & 11.17 \\
                CH 21.14 & 179.69 & 33 & - & 10.56 & CH 22.22 & 547.68 & 64 & $6.3 \pm 0.7$ & 11.05 \\
                CH 21.15 & 182.69 & 40 & $7 \pm 1^{(\ddagger)}$ & 10.21 & CH 22.23 & 548.72 & 64 & $3.5 \pm 0.5$ & 11.17 \\
                CH 21.16 & 183.66 & 47 & $5.2 \pm 0.6$ & 10.83 & CH 22.24 & 549.67 & 40 & $1.8 \pm 0.4$ & 11.53 \\
                ES 21.1 & 184.66 & 57 & $5.6 \pm 0.9^{(\dagger)}$ & 10.79 & ES 22.5 & 549.70 & 38 & $1.6 \pm 0.3$ & 11.53 \\
                CH 21.17 & 184.70 & 62 & $7.1 \pm 0.8$ & 10.79 & CH 22.25 & 550.71 & 71 & $6.5 \pm 0.7$ & 10.76 \\
                CH 21.18 & 185.64 & 55 & $10 \pm 1^{(\ddagger)}$ & 10.67 & CH 22.26 & 551.69 & 61 & $3.8 \pm 0.5$ & 11.17 \\
                CH 21.19 & 188.67 & 58 & - & 10.35 & CH 22.27 & 552.70 & 65 & $3.4 \pm 0.5$ & 11.19 \\
                ES 21.2 & 193.73 & 43 & $8 \pm 1^{(\dagger)}$ & 10.63 & CH 23.1 & 804.86 & 38 & $4.3 \pm 0.5$ & - \\
                CH 22.1 & 411.86 & 59 & $5.1 \pm 0.6$ & - & CH 23.2 & 805.88 & 57 & $3.9 \pm 0.5$ & - \\
                CH 22.2 & 418.83 & 52 & $4.6 \pm 0.5$ & - & CH 23.3 & 806.96 & 56 & $4.6 \pm 0.8^{(\ddagger)}$ & - \\
                CH 22.3 & 425.92 & 30 & $3.4 \pm 0.4$ & - & CH 23.4 & 812.89 & 41 & $1.8 \pm 0.3$ & 11.54 \\
                CH 22.4 & 434.86 & 51 & $4.5 \pm 0.5$ & - & CH 23.5 & 813.84 & 56 & $2.8 \pm 0.4$ & - \\
                CH 22.5 & 450.83 & 64 & $4.7 \pm 0.6$ & - & CH 23.6 & 814.93 & 28 & - & - \\
                CH 22.6 & 458.80 & 49 & $4.7 \pm 0.6$ & - & CH 23.7 & 815.86 & 48 & $2.4 \pm 0.5^{(\dagger)}$ & - \\
                CH 22.7 & 471.77 & 62 & $6.6 \pm 0.7$ & - & CH 23.8 & 820.83 & 53 & $2.8 \pm 0.4$ & - \\
                CH 22.8 & 525.77 & 56 & $5.5 \pm 0.6$ & - & CH 23.9 & 821.94 & 44 & $2.5 \pm 0.4$ & - \\
                CH 22.9 & 534.69 & 48 & $3.6 \pm 0.5$ & 11.16 & CH 23.10 & 831.80 & 49 & $2.5 \pm 0.6^{(\ddagger)}$ & - \\
                CH 22.10 & 535.69 & 68 & $3.5 \pm 0.7$ & 11.15 & CH 23.11 & 832.82 & 46 & $2.3 \pm 0.6^{(\ddagger)}$ & - \\
                CH 22.11 & 536.67 & 67 & $7.4 \pm 0.8^{(\dagger)}$ & 10.96 & CH 23.12 & 833.82 & 47 & $2.9 \pm 0.4$ & - \\
                ES 22.1 & 536.70 & 47 & $6.7 \pm 0.7$ & 10.96 &  &  &  &  &  \\
                \hline
        	\end{tabular}
        	\tablefoot{The MJDs were computed relative to MJD$_0 \equiv 59264.336$. The S/N was computed at 5830~{\AA}. $^{(\dagger)}$-$^{(\ddagger)}$: VF computed only with VF$_{5735}$ or VF$_{5800}$, respectively (Appendix~\ref{appendix_veiling}). The typical uncertainty on $V_{\rm mag}$ is $0.01$. The resolving power $R$ is $140000$ for the ESPRESSO spectra and $27800$ for all the CHIRON spectra except for CH 21.4, 21.14, and 21.15 which have $R = 78000$.}
        	\label{tab:log_specobs} 
        \end{table*}
    
        \section{Gaussian fits}
        \label{appendix_gaussian}
        We used a multiple Gaussian model to fit the continuum-normalized emission lines in the spectrum of RU~Lup. This model is a sum of symmetric (S) and asymmetric (A) Gaussian functions. The asymmetric Gaussian function $\mathcal{G}_{\rm{A}}$ is defined as
        \begin{equation}
            \mathcal{G}_{\rm{A}}(v) = C\exp\left[ -\frac{(v - v_0)^2}{2\sigma^2} \right] \text{, with } \sigma =
            \begin{cases}
                \sigma_{\rm b}  & \text{if $v < v_0$} \\
                \sigma_{\rm r} & \text{if $v \geq v_0$} \\
            \end{cases}
            \label{asymm_gaussian_def}
        \end{equation}
        where $C$ is the amplitude relative to the continuum, $v_0$ is the line center and $\sigma$, $\sigma_{\rm b}$, and $\sigma_{\rm r}$ are standard deviations.
        The symmetric Gaussian function $\mathcal{G}_{\rm{S}}$ has $\sigma_{\rm b} = \sigma_{\rm r}$. The model is called $n$S+$m$A in the main text, where $n$ and $m$ are the number of symmetric and asymmetric Gaussian functions, respectively. It is expressed as
        \begin{equation}
            F(v) = 1 + \sum_{i = 1}^{n} \mathcal{G}_{\rm{S}i}(v)  + \sum_{i = 1}^{m} \mathcal{G}_{\rm{A}i}(v).
        \end{equation}
 
        \section{Calculation of continuum veiling}
        \label{appendix_veiling}
        We estimated the veiling in our spectra using the ES~22.5 observation as template, which is the spectrum with the highest S/N among the ones with lower veiling in our sample. 
        This method simplifies the calculation of the veiling, because the template does not have to be rotationally broadened. In addition, there are epochs in which the line-filling emission is so strong that the fit with ROTFIT did not converge allowing $v \sin i$ to vary (e.g., ES~21.1).
        
        The veiling of the spectra relative to the ES~22.5 template (VF$_{\rm rel}$) is defined as $\rm{VF}_{\rm rel} = [F_{\rm obs}(\lambda) - F_{\rm T}(\lambda)]/F_{\rm T,c}$, where $F_{\rm obs}(\lambda)$ and $F_{\rm T}(\lambda)$ are the observed and template spectra and $F_{\rm T,c}$ is the continuum of the template.
        VF$_{\rm rel}$ can be converted to the absolute veiling; that is the veiling with respect to that of a hypothetical spectrum of  RU~Lup if it did not exhibit any accretion.
        The absolute veiling is $\rm{VF}_{\rm abs}~=~[F_{\rm obs}(\lambda) - F_{\rm 0}(\lambda)]/F_{\rm 0,c}$ where $F_{\rm 0}(\lambda)$ and $F_{\rm 0,c}$ is the wavelength-dependent flux of the non-accreting version of RU\,Lup and its continuum, respectively. 
        ${\rm VF}_{\rm abs}$ can be determined because the absolute veiling VF$_{\rm T}$ of the template could be computed independently with ROTFIT (Sect.~\ref{stellar_pars}). 
        Since VF$_{\rm T}$ is defined as $\rm{VF}_{\rm T} = [F_{\rm T}(\lambda) - F_{\rm 0}(\lambda)]/F_{\rm 0,c}$, and this relation also holds for the continua — that is, $\rm{VF}_{\rm T} = [F_{\rm T, c} - F_{\rm 0, c}]/F_{\rm 0,c}$ — it can be shown that the equation for the conversion is 
        \begin{equation}
            1 + \rm{VF_{abs}} = (1 + \rm{VF_{rel}}) \cdot (1 + \rm{VF_{T}}).
            \label{VF_rel}
        \end{equation}

        \begin{figure}
        	\centering
        	\includegraphics[width=\linewidth]{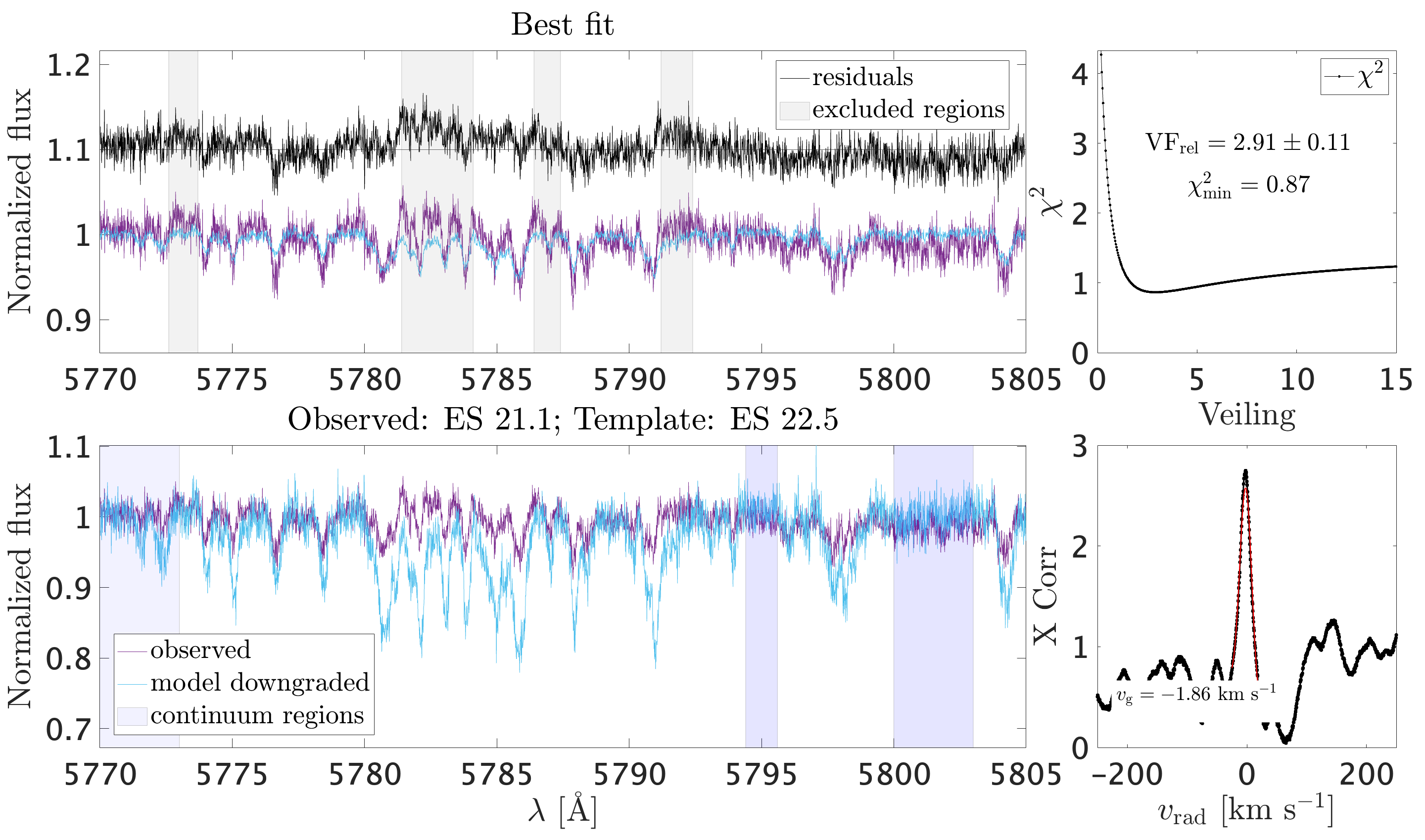}
        	\caption{Veiling calculation for the ES~21.1 spectrum. The bottom left panel shows the ES~21.1 and ES~22.5 spectra, normalized used the shaded regions. The top left panel shows the veiled ES~22.5 spectrum that best fits the ES~21.1 spectrum. The right panels display the cross correlation function (bottom) and the $\chi^2$ as a function of VF (top).}
        	\label{fig:veiling_ES21.1}
        \end{figure}

        We computed the veiling in the two spectral regions shown in Fig.~\ref{fig:ES21_vs_22_abs}. 
        % The first region is between 5725 and 5745~{\AA}, while the second is between 5770 and 5805~{\AA}. 
        We define the veiling in these two regions as VF$_{5735}$ and VF$_{5800}$. For both regions, we used VF$_{\rm T}~=~1.57 \pm 0.31$ for the ES~22.5 spectrum. 
        The procedure to estimate VF$_{\rm rel}$ consisted of the following steps:
        \begin{enumerate}
            \item normalization of the spectra with a linear fit of selected portions of the spectrum where the continuum is seen;
            \item down-grading of the template to the resolution of the observed spectrum;
            \item cross-correlation between the template and the observation to match the position of the absorption lines;
            \item exclusion of regions affected by line emission;
            \item computation of the $\chi^2$-function, defined as
            \begin{equation}
                \chi^2 (\rm{VF_{\rm rel}}) = \sum_{\lambda_i} \left( F_{\rm obs}(\lambda_i) -  \frac{F_{\rm T}(\lambda_i) + \rm{VF}_{\rm rel}}{1 + \rm{VF}_{\rm rel}} \right)^2
            \end{equation}
            as a function of VF$_{\rm rel}$.
        \end{enumerate}
        The best fitting VF$_{\rm rel}$ was found minimizing the $\chi^2$, and the uncertainty as the standard deviation of $\exp(-\chi^2)$. Then, we converted VF$_{\rm rel}$ to VF$_{\rm abs}$ using Eq.~\ref{VF_rel}.
        
        Figure~\ref{fig:veiling_ES21.1} shows the procedure to compute VF$_{5800}$ for the ES~21.1 spectrum, pointing out the effect of line emission in the estimation of the VF. 
        Due to the resolution, the S/N, and the effect of line emission, sometimes the fits did not converge. Therefore, we selected the observations for which VF$_{5735}$ and VF$_{5800}$ differed by more than 1$\sigma$, we visually inspected the results of the procedure for both regions, and we excluded the cases that we considered unclear. When both VF$_{5735}$ and VF$_{5800}$ agreed with each other, we derived a single measure of VF$_{\rm abs}$ by computing a weighted average (with weights $w_i = 1/d\rm{VF}_i^2$) of the two values. The results are reported in Table~\ref{tab:log_specobs}.

        \section{Spot model}
        \label{spot_model}
        \begin{figure*}
        	\centering
        	\includegraphics[width=\linewidth]{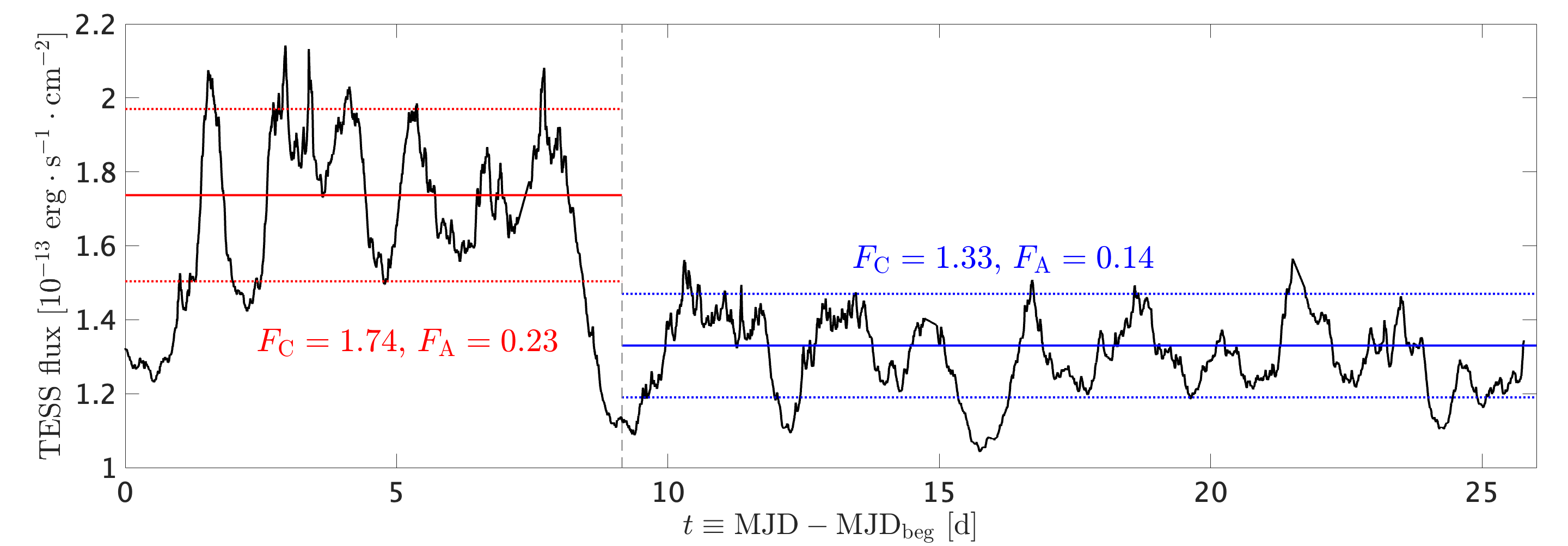}
        	\caption{TESS Sector~65 light curve converted in absolute flux in the TESS bandpass. The solid lines mark the zero point of the oscillations, while the dotted lines mark the maximum and the minimum of the oscillations. The vertical dashed lines indicate the transition between the two portions. $F_{\rm C}$ and $F_{\rm A}$ are the zero point and the amplitude fluxes, respectively, in units of $10^{-13} ~ \rm{erg~s^{-1}~cm^{-2}}$.}
        	\label{fig:TESS_2023_spot_analysis}
        \end{figure*}
        We derived an analytical model to reproduce the variability of the TESS light curve.
        We assumed a spot with a constant temperature $T$, filling factor $f$, located at a latitude $\theta_{\rm S}$ and azimuth $\phi$ on the surface of the star. The spot radiates as a black-body: its energy per unit time, area, solid angle, and wavelength interval is given by
        \begin{equation}
            B_{\lambda}(T) = \frac{2hc^5}{\lambda^5} \frac{1}{\exp{(hc/k_{\rm B}T)} - 1},
        \end{equation}
        where $h$ is the Plank constant, $c$ is the speed of light, $k_{\rm B}$ is the Boltzmann constant, and $\lambda$ is the wavelength.
        The projected surface of the spot varies as
        \begin{equation}
            S(\phi) = S_0 (\hat{n} \cdot \hat{s}),
        \end{equation}
        where $S_0 = 4\pi R_{\star}^2 f$ is the surface of the spot, $\hat{n}~=~(\cos\theta_{\rm S} \cos \phi, \cos \theta_{\rm S} \sin \phi, \sin \theta_{\rm S})$ is the normal to the surface, and $\hat{s} = (\sin i, 0, \cos i)$ is the line of sight toward an observer located at an angle $i$ from the rotation axis.
        Therefore, the flux observed with TESS is
        \begin{multline}
             F(T, f, \theta_{\rm S}, \phi) = F_{\star} + \left[ \frac{R_{\star}^2}{d^2} f \int B_{\lambda}(T) R(\lambda) d\lambda \right] \\
             \cdot (\cos\theta_{\rm S} \sin i \cos \phi + \sin \theta_{\rm S} \cos i)
        \end{multline}
        where $d$ is the distance to the system, $R(\lambda)$ is the response function of the TESS filter, and 
        \begin{equation}
             F_{\star} = \frac{R_{\star}^2}{2d^2} \int B_{\lambda}(T_{\rm eff}) R(\lambda) d\lambda = 2.7 \times 10^{-14} ~ \rm{erg~s^{-1}~cm^{-2}}
        \end{equation}
        is the integrated flux of the star in the TESS bandpass\footnote{The factor 1/2 comes from the fact that we see only one hemisphere.}. We downloaded the TESS response function from the SVO Filter Profile Service\footnote{\url{http://svo2.cab.inta-csic.es/theory/fps/}} \citep{SVO_2020}.
        The modulation can be written as
        \begin{equation}
             F(T, f, \theta_S, \phi) = F_{\star} + A \cos \phi + C.
        \end{equation}
        with
        \begin{equation}
            A = \left[ \frac{R_{\star}^2}{d^2} f \int B_{\lambda}(T) R(\lambda) d\lambda \right] \cos\theta_{\rm S} \sin i  
        \end{equation}
        and
        \begin{equation}
            C = A \frac{\tan \theta_{\rm S}}{\tan i}.
            \label{spot_latitude}
        \end{equation}
        The latitude of the spot can be directly obtained from the inversion of Eq.~\ref{spot_latitude}, while from the amplitude $A$ of the oscillations we get a relation between $f$ and $T$
        \begin{equation}
            f = A \frac{d^2}{R_{\star}^2} \left[ \int B_{\lambda}(T) R(\lambda) d\lambda \right]^{-1} (\cos\theta_{\rm S} \sin i)^{-1}.
            \label{spot_fvsT}
        \end{equation}

        We converted the TESS magnitude $T$ to flux $F$ using the formula $F = \rm{ZP}_{\lambda}~10^{-T/2.5}$, where $\rm{ZP}_{\lambda} ~ = ~ 1.33 ~ \times ~ 10^{-9} ~ \rm{erg~s^{-1}~cm^{-2}}$ is the zero point flux of TESS, that we took from the SVO Filter Profile Service. Then, we visually estimated the amplitude $F_{\rm A}$ and the zero point $F_{\rm C}$ of the oscillations for the two portions of the light curve, defined as in Sect.~\ref{TESS}. The values are reported in Fig.~\ref{fig:TESS_2023_spot_analysis}.
        The latitude of the spot and the relation between $f$ and $T$ can be derived from Eqs.~\ref{spot_latitude} and \ref{spot_fvsT} with $C = F_{\rm C} - F_{\star}$ and $A = F_{\rm A}$.  We derived $\theta_{\rm{S}1} = 61^{\rm o}$ and $\theta_{\rm{S}2} = 65^{\rm o}$ for the latitude of the spot in the two segments. The relation between $f$ and $T$ (Eq.~\ref{spot_fvsT}) is shown in Fig.~\ref{fig:fvsT_spot}. 

        % \textcolor{blue}{The amount of veiling as a function of $\lambda$ produced by the spot can be estimated in a similar way. The procedure to obtain the spectrum emitted by the spot is the same as what we outlined to obtain the flux in the TESS bandpass, but without the integration over the response function. We get
        % \begin{multline}
        %     F_{\rm S}(\lambda, \phi) = \frac{R_{\star}^2}{d^2} f \cdot B_{\lambda}(T) \\
        %     \cdot (\cos\theta_{\rm S} \sin i \cos \phi + \sin \theta_{\rm S} \cos i).
        % \end{multline}
        % The stellar spectrum is instead
        % \begin{equation}
        %    F_{\rm \star}(\lambda) = \frac{R_{\star}^2}{2d^2} f \cdot B_{\lambda}(T_{\rm eff}).
        % \end{equation}
        % Therefore, we derive the veiling fraction $\rm{VF}_{\lambda}(\phi) = F_{\rm S}(\lambda, \phi)/F_{\star}(\lambda)$ as
        %\begin{equation}
        %    \rm{VF}_{\lambda} (\phi) = 2f \frac{B_{\lambda}(T)}{B_{\lambda}(T_{\rm eff})} \cdot (\cos\theta_{\rm S} \sin i \cos \phi + \sin \theta_{\rm S} \cos i)
        %\end{equation}
        % }

        \section{Filling factor of a spherical cap}
        \label{f_cap}
        
        A spherical cap is a portion of a sphere cut by a plane. In spherical coordinates, it is defined as
        \begin{equation}
            \big\{ (\theta, \phi) \quad | \quad \frac{\pi}{2} - \Delta \theta_{\rm S} \leq \theta \leq \frac{\pi}{2}, \quad 0 \leq \phi \leq 2\pi \big\}
        \end{equation}
        where $\theta$ and $\phi$ are the latitude and the azimuth, and $\Delta \theta_{\rm S}$ is the half-aperture angle of the cap (spot). The filling factor of the cap can be computed by integrating
        \begin{equation}
            f = \frac{1}{4 \pi R_{\star}^2} \int_{\pi/2 - \Delta \theta_{\rm S}}^{\pi/2} \int_{0}^{2 \pi} R_{\star}^2 \cos \theta d\theta d\phi = \frac{1}{2} (1 - \cos \Delta \theta_{\rm S}).
        \end{equation}
        Inverting this relation, we obtain
        \begin{equation}
            \Delta \theta_{\rm S} = \arccos{(1 - 2f)}.
        \end{equation}

    \end{appendix}
	
\end{document}